\documentclass{aastex62}

\usepackage{hyperref}
\newcommand\blfootnote[1]{%
	\begingroup
	\renewcommand\thefootnote{}\footnote{#1}%
	\addtocounter{footnote}{-1}%
	\endgroup
}

\graphicspath{{./}{figures/}}

\shortauthors{K. Barkaoui et al.}

\begin{document}
\pdfoutput=1 
\title{Discovery of three new transiting hot-Jupiters: WASP-161~b, WASP-163~b and WASP-170~b}

\correspondingauthor{K. Barkaoui}
\email{khalid.barkaoui@doct.uliege.be}

\author{K. Barkaoui}
\affiliation{Space sciences, Technologies and Astrophysics Research (STAR) Institute, Universit\'e de Li\`ege, Belgium}
\affiliation{Oukaimeden Observatory, High Energy Physics and Astrophysics Laboratory, Cadi Ayyad University, Marrakech, Morocco}

\author{A. Burdanov}
\affiliation{Space sciences, Technologies and Astrophysics Research (STAR) Institute, Universit\'e de Li\`ege, Belgium}

\author{C. Hellier}
\affiliation{Astrophysics Group, Keele University, Staffordshire, ST5 5BG, UK}

\author{M. Gillon} 
\affiliation{Space sciences, Technologies and Astrophysics Research (STAR) Institute, Universit\'e de Li\`ege, Belgium}

\author{B. Smalley}
\affiliation{Astrophysics Group, Keele University, Staffordshire, ST5 5BG, UK}

\author{P. F. L. Maxted}
\affiliation{Astrophysics Group, Keele University, Staffordshire, ST5 5BG, UK}

\author{M. Lendl}
\affiliation{Space Research Institute, Austrian Academy of Sciences, Schmiedlstr. 6, 8042 Graz, Austria}
\affiliation{Observatoire astronomique de l'Universit\'e de Geneve, 51 ch. des Maillettes, 1290 Sauverny, Switzerland}

\author{A. H. M. J. Triaud}
\affiliation{School of Physics \& Astronomy, University of Birmingham, Edgbaston, Birmingham B15 2TT, United Kingdom}

\author{D. R. Anderson}
\affiliation{Astrophysics Group, Keele University, Staffordshire, ST5 5BG, UK}

\author{J. McCormac}
\affiliation{ Department of Physics, University of Warwick, Gibbet Hill Road, Coventry, CV4 7AL, UK}

\author{E. Jehin}
\affiliation{Space sciences, Technologies and Astrophysics Research (STAR) Institute, Universit\'e de Li\`ege, Belgium}

\author{Y. Almleaky}
\affiliation{Space and Astronomy Department, Faculty of Science, King Abdulaziz University, 21589 Jeddah, Saudi Arabia}
\affiliation{King Abdullah Centre for Crescent Observations and Astronomy, Makkah Clock, Mecca 24231, Saudi Arabia}

\author{D. J. Armstrong}
\affiliation{ Department of Physics, University of Warwick, Gibbet Hill Road, Coventry, CV4 7AL, UK}

\author{Z. Benkhaldoun}
\affiliation{Oukaimeden Observatory, High Energy Physics and Astrophysics Laboratory, Cadi Ayyad University, Marrakech, Morocco}

\author{F. Bouchy}
\affiliation{Observatoire astronomique de l'Universit\'e de Geneve, 51 ch. des Maillettes, 1290 Sauverny, Switzerland}

\author{D. J. A. Brown}
\affiliation{ Department of Physics, University of Warwick, Gibbet Hill Road, Coventry, CV4 7AL, UK}

\author{A. C. Cameron}
\affiliation{School of Physics and Astronomy, University of St Andrews, North Haugh, St Andrews, Fife KY16 9SS}

\author{A. Daassou}
\affiliation{Oukaimeden Observatory, High Energy Physics and Astrophysics Laboratory, Cadi Ayyad University, Marrakech, Morocco}

\author{L. Delrez}
\affiliation{Cavendish  Laboratory, J J Thomson Avenue, Cambridge, CB3 0HE, UK}
\affiliation{Space sciences, Technologies and Astrophysics Research (STAR) Institute, Universit\'e de Li\`ege, Belgium}

\author{E. Ducrot}
\affiliation{Space sciences, Technologies and Astrophysics Research (STAR) Institute, Universit\'e de Li\`ege, Belgium}

\author{E. Foxell}
\affiliation{ Department of Physics, University of Warwick, Gibbet Hill Road, Coventry, CV4 7AL, UK}

\author{C. Murray}
\affiliation{Cavendish  Laboratory, J J Thomson Avenue, Cambridge, CB3 0HE, UK}

\author{L. D. Nielsen}
\affiliation{Observatoire astronomique de l'Universit\'e de Geneve, 51 ch. des Maillettes, 1290 Sauverny, Switzerland}

\author{F. Pepe}
\affiliation{Observatoire astronomique de l'Universit\'e de Geneve, 51 ch. des Maillettes, 1290 Sauverny, Switzerland}

\author{D. Pollacco}
\affiliation{ Department of Physics, University of Warwick, Gibbet Hill Road, Coventry, CV4 7AL, UK}

\author{F. J. Pozuelos}
\affiliation{Space sciences, Technologies and Astrophysics Research (STAR) Institute, Universit\'e de Li\`ege, Belgium}

\author{D. Queloz}
\affiliation{Cavendish  Laboratory, J J Thomson Avenue, Cambridge, CB3 0HE, UK}
\affiliation{Observatoire astronomique de l'Universit\'e de Geneve, 51 ch. des Maillettes, 1290 Sauverny, Switzerland}

\author{D. Segransan}
\affiliation{Observatoire astronomique de l'Universit\'e de Geneve, 51 ch. des Maillettes, 1290 Sauverny, Switzerland}

\author{S. Udry}
\affiliation{Observatoire astronomique de l'Universit\'e de Geneve, 51 ch. des Maillettes, 1290 Sauverny, Switzerland}

\author{S. Thompson}
\affiliation{Cavendish  Laboratory, J J Thomson Avenue, Cambridge, CB3 0HE, UK}

\author{R. G. West}
\affiliation{ Department of Physics, University of Warwick, Gibbet Hill Road, Coventry, CV4 7AL, UK}



\begin{abstract}

We present the new discovery  of three new transiting hot-Jupiters by the WASP-South project, WASP-161~b, WASP-163~b and WASP-170~b. Follow-up radial velocities obtained with the Euler/CORALIE spectrograph and transit light-curves obtained with the TRAPPIST-North, TRAPPIST-South, SPECULOOS-South, NITES, and Euler telescopes have enabled us to determine the masses and radii for these transiting exoplanets.  WASP-161\,b completes an orbit around its $V=11.1$ F6V-type host star in 5.406 days, and has a mass $M_p = 2.5\pm 0.2$$M_{Jup}$ and radius  $R_p = 1.14\pm 0.06$ $R_{Jup}$.  WASP-163\,b orbiting around its host star (spectral type G8V and the magnitude  $V=12.5$ )  each 1.609 days, and has a mass  $M_P = 1.9\pm0.2$ $M_{Jup}$ and a radius $R_p = 1.2\pm0.1$ $R_{Jup}$. WASP-170\,b has a mass of $1.7\pm0.2$ $M_{Jup}$ and a radius of $1.14\pm0.09$ $R_{Jup}$, is on a 2.344 days orbit around a G1V-type star of magnitude $V=12.8$. 
Given their irradiations ($\sim10^9$ erg.s$^{-1}$.cm$^{-2}$) and masses, the three new planets' sizes are in the good agreement with classical  models of irradiated giant planets.

\end{abstract} 



\section{Introduction} \label{sec:intro}

Inaugurated by the seminal discovery of 51 Peg b in 1995 \citep{mayor1995}, the study of exoplanets has  dramatically developed to become one of the most important fields of modern astronomy.
Since 1995,  the number of exoplanets detected, most of them by the transit technique  \citep{charbonneau2000, Henry2000}.  

Among this large harvest, highly irradiated giant planets (aka hot Jupiters) transiting bright nearby stars have a particular scientific interest. These rare objects - $<1$\% of solar-type stars \citep{Winn2015} - undergo irradiation orders of magnitude larger than any solar system planets \citep{Fortney2007}, and are also subject to intense gravitational and magnetic fields \citep{Correia2010, Chang2010}. Studying in detail their physical and chemical response to such extreme conditions provides a unique opportunity to improve our knowledge on planetary structure, composition and physics. The brightness of their host star combined to their eclipsing configuration makes possible such detailed characterization, notably to measure precisely their size, mass, and  orbital parameters  \citep{Winn2010, Deming2009}, but also to probe their atmospheric properties, for example the $P-T$ profiles, chemical composition and albedos  \citep{Seager2010, Sing2016, Crossfield2015}. 

The WASP (Wide Angle Search for Planets) project (described in \cite{Pollacco2006, Cameron2007} )uses two robotic installations, one at La Palma (Spain) and one at Sutherland (South Africa), to scout the sky for gas giants transiting the solar type stars. With more than 100  hot Jupiters discovered so far in front of bright nearby stars, WASP is a key contributor to the  study of highly irradiated giant planets.  In this paper, we report the discovery   of three new gas giants, WASP-161\,b, WASP-163\,b and WASP-170\,b, transiting bright ($V$= 11.1, 12.5 \& 12.8) solar-type (F6-, G8- and G1-type) dwarf stars. 

In Section \ref{sec:obseva_and_data}, we present the observations used to discover WASP-161\,b, WASP-163\,b and WASP-170\,b, and to confirm their planetary natures and measure their parameters. In Section \ref{sec:TN}, we describe notably TRAPPIST-North, a 60cm robotic telescope installed recently by the University of Li\`ege at Oukaimeden observatory (Morocco), that played a significant role in the confirmation and characterization of the planets. Section \ref{sec:spectro_analysis} presents the determination of the atmospheric parameters of the host stars. In Section \ref{sec:mcmc}, we describe our global analysis of the dataset for the three planetary systems that enabled us to determine their main physical and orbital parameters. We discuss briefly our results in Section \ref{sec:conclusion}.

\section{Observations and data reduction} \label{sec:obseva_and_data}
\subsection{WASP photometry} \label{sec:wasp}
\begin{figure}[h!]
	\centering
	\includegraphics[scale=0.34]{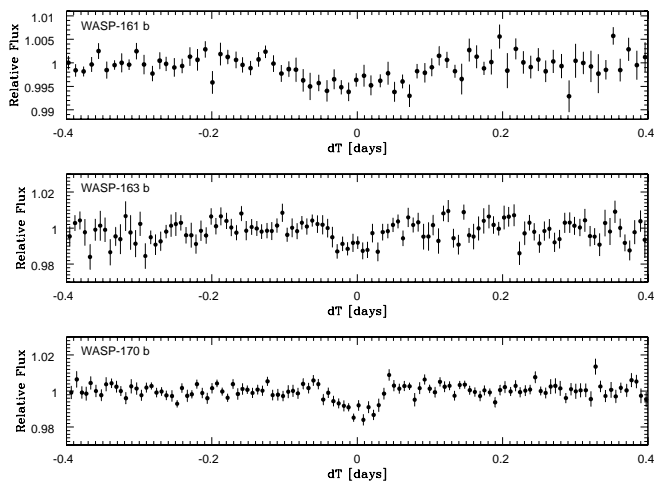}
	\caption{The light curve of WASP-161 ({\it top}), WASP-163 ({\it middle}) and WASP-170 ({\it bottom}) (binned = 10 min)  folded on the  transit ephemeris from the transit search algorithm described in \cite{Cameron2006}.}
	\label{LCsWASP}
\end{figure}

WASP-161 and WASP-170 (see Table \ref{mcmcwasp161163170} for coordinates and magnitudes) were observed with the WASP-South \citep{Hellier2011,Hellier2012} in 2011 and 2012, while WASP-163 was observed in 2010 and 2012. The WASP-South data reduction  methods described by \cite{Cameron2006}, and selected \citep{Collier2007} as valuable candidates showing possible transits of short-period ($\sim$ 5.4, 1.6, and 2.3 days) planetary sizes bodies (Fig. \ref{LCsWASP}).

\begin{table*}
	
	\begin{center}
		{\renewcommand{\arraystretch}{1.2}
			\resizebox{0.97\textwidth}{!}{
				\begin{tabular}{lcccc}
					\hline
					& \multicolumn{3}{c}{Star general informations }  \\
					\hline
					& WASP-161  & WASP-163  & WASP-170   \\
					& 2MASS08252108-1130035 & 2MASS17060901-1024467 & 2MASS09013992-2043133 \\
					& GaiaId 5751177091580191360 & GaiaId 4334991786994866304 & GaiaId 5656184406542140032\\
					\hline
					RA (J200)     &   $ 08^h 25^m 21.09^s  $  & $ 17^h 06^m 08.98^s  $   &  $ 09^h 01^m 39.93^s  $   \\
					Dec (J200)   &  $ -11^\circ 30' 03.6'' $  & $ -10^\circ 24' 47.0'' $  &  $ -20^\circ 43' 13.6'' $ \\
					Vmag [UCAC4]   &    10.98    & 12.54  &  12.65  \\
					Jmag [2MASS]   &     10.09   &  10.67  & 11.13 \\
					Gmag [Gaia-DR1] & 10.84 & 12.13 & 12.36 \\
					Parallax [mas] [$Gaia$-DR2]  &  $ 2.8864 \pm 0.0345$   &  $ 3.7981 \pm 0.0525$  &  $ 3.2439 \pm 0.0390$   \\
					\hline
					&    \multicolumn{3}{c}{Stellar parameters from spectroscopic analysis }  \\
					\hline
					$T_{eff}$ (K)  & $6400\pm 100$  &  $5500\pm 200$   &  $5600\pm 150$   \\
					$\log g_\star$ [cgs] &  $4.5 \pm 0.15$  &  $4.0 \pm 0.3$   &   $4.0 \pm 0.2$  \\
					$[Fe/H]$  & $+0.16\pm 0.09$ &  $-0.34\pm 0.21$  & $+0.22\pm 0.09$ \\
					Spectral type   & F6 &  G8 & G1 \\
					$V \sin i$ [Km/s]  & $18\pm 0.8$ &  $< 5$ & $5.6\pm 1$ \\
					$\log A(Li)$ & No Lithium seen & $< 1.6$  & $1.52 \pm 0.09 $ \\

					\hline
					&  \multicolumn{3}{c}{Parameters from MCMC analysis}   \\
					\hline
					MCMC Jump parameters      &  &    \\
					\hline
					 $(R_p/R_\star)^2$ [\%]        &    $0.45092 \pm 0.00023$                  &   $1.417 \pm 0.067$              &  $1.382 \pm 0.001$ \\   
					Impact parameter $b$  $[R_\star]$                   & $0.14 _{-0.10}^{+0.15}$             &  $0.45 ^{+0.09}_{-0.06}$       &   $0.689 \pm 0.021$ \\
					Transit duration $W$ [d]                                     & $0.2137 \pm 0.0022$  &  $0.093 \pm 0.001$                &  $0.085 \pm 0.001$   \\
					Mid-transit $T_0$ [HJD]                                                        & $7416.5289 \pm 0.0011$            &   $7918.4620 \pm 0.0004$      &   $7802.3915 \pm 0.0002$   \\
					Orbital period $P$ [d]                                          &  $ 5.4060425 \pm 0.0000048$     &  $1.6096884 \pm 0.0000015$ &  $2.34478022 \pm 0.0000036$  \\ 
					RV $K_2$ $[\text{m.s}^{-1}\text{d}^{1/3}]$           & $405 \pm 20$                              &  $386.69 \pm 16$                     &  $340 \pm 20$  \\
					Effective temperature $T_{eff}$ [K]                     & $6406 \pm 100$                           & $5499  \pm 200$                      &  $5593 \pm 150$  \\
					Metallicity $[Fe/H]$                                              & $0.16 \pm 0.09$                          &  $-0.34 \pm 0.21$                      &  $0.21 \pm 0.19$   \\
					\hline
					Deduced stellar parameters from MCMC  & & \\
					\hline
					Mean density $\rho_\star$ [$\rho_\odot$]                                   & $0.282_{-0.027}^{+0.013}$                &  $0.92 _{-0.10}^{+0.13}$             &  $1.121 _{-0.076}^{+0.093}$    \\
					Stellar surface gravity  $\log g_\star$ [cgs]                 & $4.111 _{-0.033}^{+0.023}$          &  $4.411 ^{+0.042}_{-0.040}$        &  $4.466 \pm 0.031$    \\
					Stellar mass     $M_\star$ [$M_\odot$]               & $1.39 \pm 0.14$                              &  $0.97 \pm 0.15$                          & $0.93 \pm 0.15$   \\
					Stellar radius     $R_\star$ [$R_\odot$]              & $1.712_{-0.072}^{+0.083}$                  &  $1.015 ^{+0.071}_{-0.074}$          &  $0.938 ^{+0.056}_{-0.061}$  \\
					Luminosity $L_\star$ [$L_\odot$]                      & $4.44 _{-0.48}^{+0.56}$                   & $0.84 ^{+0.20}_{-0.17}$                &  $0.77 \pm 0.14$  \\
					\hline
					Deduced planet parameters from MCMC  & WASP-161 b  & WASP-163 b  & WASP-170 b   \\
					\hline
					RV $K$ [m$\text{s}^{-1}$]                                & $ 230 \pm 12$                                     & $329 \pm 14$                             & $ 255 \pm 15$   \\ 
					Planet/star radius ratio  $R_p/R_\star$               & $0.0671 \pm 0.0017$                             &  $0.119 \pm 0.003$                          & $0.1175 \pm 0.0041$  \\  
					Impact parameter $b$ $[R_\star]$                  & $0.14 _{-0.10}^{+0.15}$                        &  $0.448 ^{+0.063}_{-0.094}$             &  $0.689 \pm 0.021$  \\
					Semi-major axis $a/R_\star$               & $8.49 ^{+0.13}_{-0.28}$                         &   $5.62 ^{+0.26}_{-0.21}$                    & $7.71 ^{+0.21}_{-0.18}$   \\
					Orbital semi-major axis $a$ [AU]                 &  $ 0.0673 \pm 0.0023$                            &  $ 0.0266 \pm 0.0014$                       &   $ 0.0337 \pm 0.0018$  \\
					 Inclination $i_p$ [deg]                     & $ 89.01 _{-1.0}^{+0.69}$                              &   $ 85.42 ^{+1.10}_{-0.85}$                             &  $ 84.87 \pm 0.28$  \\
					Density $\rho_p$ [$\rho_{Jup}$]                & $ 1.66 \pm 0.22$                         & $ 1.07  ^{+0.23}_{-0.17}$                       & $ 1.21  ^{+0.24}_{-0.19}$  \\
					Surface gravity $\log g_p$ [cgs]               & $ 3.69 _{-0.42}^{+0.37}$                          & $ 3.52 \pm 0.05$                                   &  $ 3.54 \pm 0.05$  \\
					Mass $M_p$ [$M_{Jup}$]                          & $ 2.49 \pm 0.21$                                    &   $ 1.87 \pm 0.21$                &   $ 1.6 \pm 0.2$  \\
					Radius $R_p$ [$R_{Jup}$]                         & $ 1.143_{-0.058}^{+0.065}$                          & $ 1.202 \pm 0.097$                                 &  $ 1.096 \pm 0.085$   \\
					Roche limit $a_R$ [AU]                            & $ 0.01101 ^{+0.00075}_{-0.00068} $                              &  $ 0.011 \pm 0.001 $                                &   $ 0.011 \pm 0.001 $  \\
					$a/a_R$                                                   &  $ 6.12 _{-0.28}^{+0.25}$                            & $ 2.35 ^{+0.16}_{-0.13}$                          &  $ 3.15 \pm 0.19$  \\
					Equilibrium temperature $T_{eq}$ [K]         & $ 1557 _{-29}^{34}$                               & $ 1638 \pm 68$                                        & $ 1422 \pm 42$   \\
					Irradiation [$\text{erg} \text{.s}^{-1} \text{.cm}^{-2}$]   & $1.35^{+0.34}_{-0.26} \times 10^9 $     & $1.63\pm 0.45 \times 10^9 $     &  $9.3 ^{+2.3}_{-2.5}\times 10^8 $    \\
					\hline
		\end{tabular}}}
	\end{center}
	\caption{The parameters of the WASP-161, WASP-163, and WASP-170 planetary systems (values + 1$\sigma$ error bars), as deduced from our data analysis presented in Section \ref{sec:analysis} }
	\label{mcmcwasp161163170}
\end{table*}

\subsection{Follow-up Photometry} \label{sec:photometry}

\subsubsection{TRAPPIST-North} \label{sec:TN}

TRAPPIST-North is a new robotic telescope of 60-cm diameter installed in June 2016 at  Oukaimeden Observatory (Morocco). It is installed by the University of Li\`ege (Belgium) and  in collaboration with the Cadi Ayyad University of Marrakech (Morocco). TRAPPIST-North extends the TRAPPIST project to the Northern hemisphere, and, as its Southern twin TRAPPIST-South, aims to the detection and characterization of transiting exoplanets, and the study of comets and other small bodies (e.g. asteroids) in the Solar System. 
The exoplanet program of TRAPPIST (75\% of its observational time) is dedicated to several programs: participating to the SPECULOOS project that aims to explore the nearest ultracool dwarf stars for transiting terrestrial planets \citep{Gillon2017Natur,Gillon2018,Burdanov2017,Delrez2018};  the search for the transit of planets previously detected by radial velocity \citep{Bonfils2011}; the follow-up of transiting planets of high interest (e.g. \citealt{Gillon2012}); and the follow-up of transiting planet candidates identified by wide-field transit surveys like WASP (e.g. \citealt{Delrez2014}). TRAPPIST-North has a F/8 Ritchey-Chretien optical design and protected by a 4.2 meters diameter dome equipped with a weather station and independent rain and light sensors. The telescope is equipped with a thermoelectrically-cooled 2048$\times$2048 deep-depletion Andor IKONL BEX2 DD CCD camera that has a pixel scale of 0.60'' that translates into a FOV of 19.8' $\times$ 19.8'. It is coupled to a direct-drive mount of German equatorial design.  For more technical details and performances of the TRAPPIST telescopes described in \cite{Jehin2011}.

TRAPPIST-North observed two partial transits of WASP-161\,b in the Sloan-$z'$ filter (20 Dec 2017 and 12 Feb 2018),  two partial transits  and one full transit of WASP-163 b in the $I+z$ filter (24 Apr, 02 May, and 13 Jun 2017),   and three partial transits of WASP-170 b in the $I+z$ (19 Apr 2017 and 11 Jan 2018) and $V$ (17 Feb 2017) filters. The reduction and photometric analysis of the data were performed as described in \cite{Gillon2013}. The resulting light curves are shown in figures \ref{wasp161lc}, \ref{wasp163lc}, and \ref{wasp170lc}.

\subsubsection{TRAPPIST-South} \label{sec:TS}
We used the 60cm robotic telescope TRAPPIST-South (TRansiting Planets and PlanetesImals Small Telescope; \citealt{Gillon2011,Jehin2011}) at La Silla (Chile) to observe a partial transit of WASP-161\,b in the Sloan-$z'$ filter on 28 Jan 2016, two partial transits of WASP-163\,b in a broad $I+z$ filter   on 6 Sep 2014 and 5 July 2016,  and two transits (one full + one partial) of WASP-170\,b in $I+z$ on 25 Dec 2015 and 26 Feb 2017. 
TRAPPIST-South is equipped with a thermoelectrically-cooled 2K $\times$ 2K CCD with the pixel scale of 0.65'' that translates into a 22' $\times$ 22' of FOV. Standard calibration of the images, fluxes extraction and differential photometry were then performed as described in \cite{Gillon2013}. The resulting light curves are shown in figures \ref{wasp161lc}, \ref{wasp163lc}, and \ref{wasp170lc}.

\subsubsection{EulerCam} \label{sec:EulerCam}
We used the EulerCam camera \citep{Lendl2012} on the 1.2-m Euler-Swiss telescope  at La Silla Observatory in Chile to observe a transit of WASP-163\,b on 27 July 2016 in the RG filter, and also a transit of WASP-170\,b on 20 Dec 2016  in the broad NGTS filter ($\lambda_{NGTS} = [500 - 900nm]$, \citealt{Wheatley2017}).   The calibration and photometric reduction (aperture + differential photometry) of the images were performed as described by \cite{Lendl2012}. The resulting light curves are shown in figures \ref{wasp163lc} and \ref{wasp170lc}.

\subsubsection{NITES} \label{sec:NITES}
We use 0.4-m NITES (Near-Infrared Transiting ExoplanetS Telescope, \citealt{McCormac2014}) robotic telescope at La Palma (Canary Islands) to observe two transits of WASP-163 b. The first transit was full and observed in R-band on 27 June 2016, while the second was only partial and observed in I-band on 10 July 2016. NITES is equipped with a 1024$\times$1024 CCD camera that has a pixel scale of 0.66$\arcsec$ that translates into a FOV of 11.3' $\times$ 11.3'. The standard calibration of the science images, fluxes extraction and differential photometry were then performed as described in \cite{Craig15,Barbary16,Bertin96,McCormac2013}. The resulting light curves are shown in figure \ref{wasp163lc}.

\subsubsection{SPECULOOS-South} \label{sec:SPC}
We use 1-m robotic SSO-Europa telescope, one of the four telescopes of the SPECULOOS-South facility, (more details found in \cite{Delrez2018,Gillon2018,Burdanov2017}) to observe one full-transit of WASP-161~b on 5 Jan 2018 in the Sloan-$z'$ filter. Each 1-m robotic telescope  equipped with 2K$\times$2K CCD camera, with good sensitivities in the very-near-infrared up to 1 $\mu$m. The calibration and photometric reduction of the data were performed as described in \cite{Gillon2013}. The resulting light curve is shown in figure \ref{wasp161lc}.

\subsection{Follow-up spectroscopy} \label{sec:spectroscopy}
\begin{figure*}
	\begin{center}
		\includegraphics[scale=0.7]{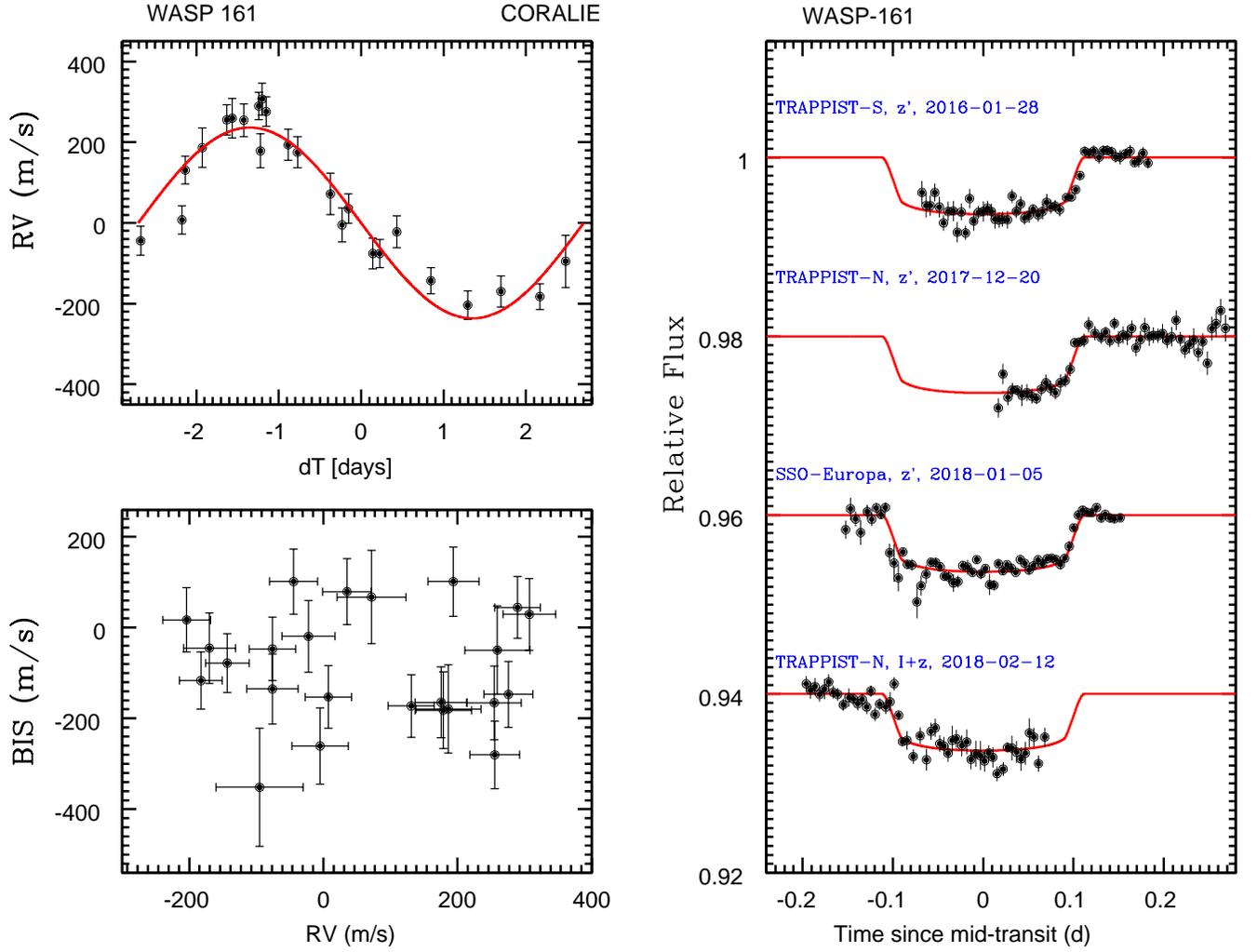} 
	\end{center}
	\caption{{\it Right-hand panel}: Individual follow-up transit light curves for WASP-161 binned per 0.005d (7.2min).  The solid red lines are the best-fit transit models. We shifted the light curves  along the $y$ axis for clarity. {\it Left-hand panel, top}: CORALIE RVs for WASP-161 with the best-fit Keplerian model in red. {\it Left-hand panel, bottom}: bisector spans (BIS) $vs$ RVs diagram.}
	\label{wasp161lc}
\end{figure*}

\begin{figure*}
	\begin{center}
		\includegraphics[scale=0.7]{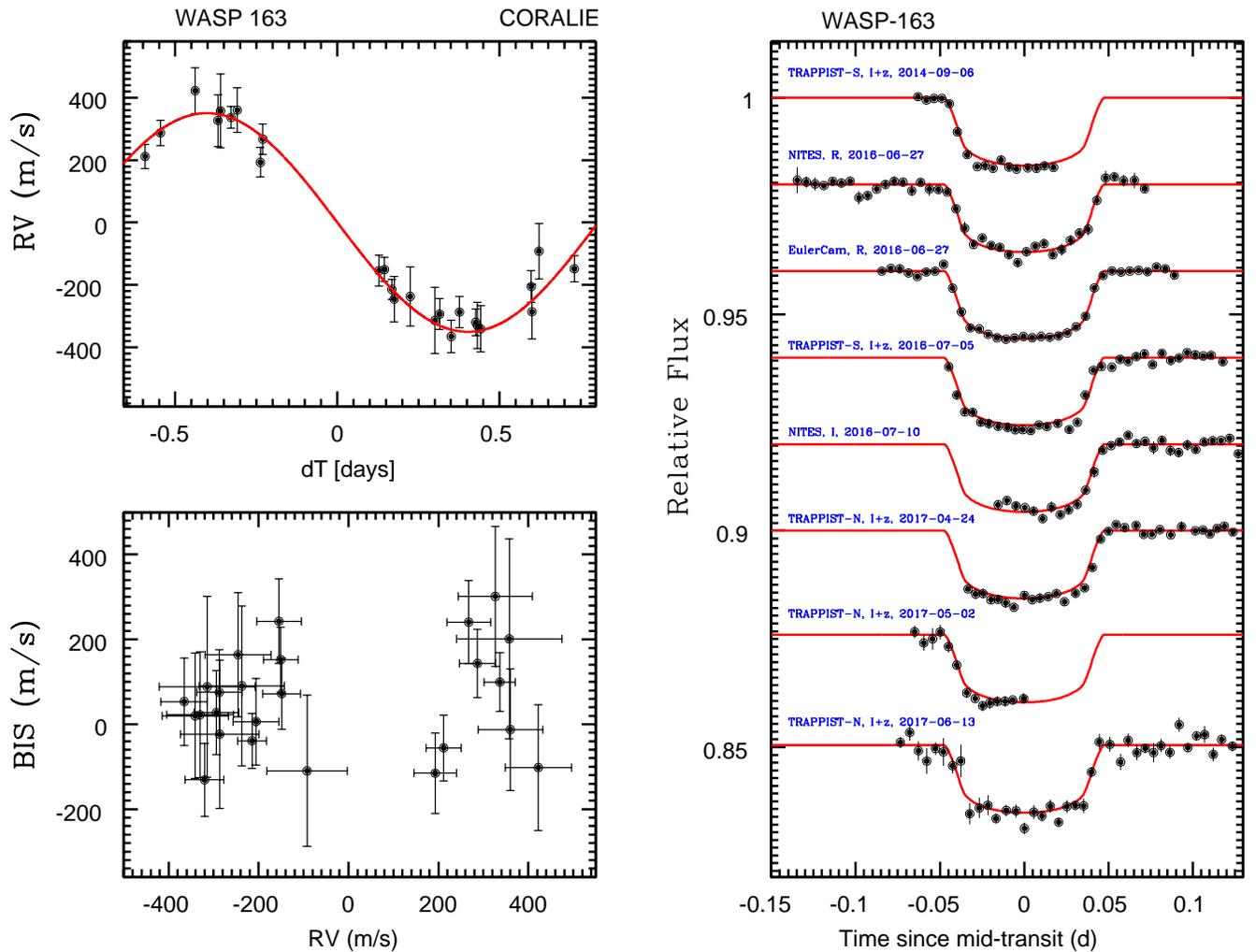} 
	\end{center}
	\caption{Same as Fig. \ref{wasp161lc} but for WASP-163.}
	\label{wasp163lc}
\end{figure*}

\begin{figure*}
	\begin{center}
		\includegraphics[scale=0.7]{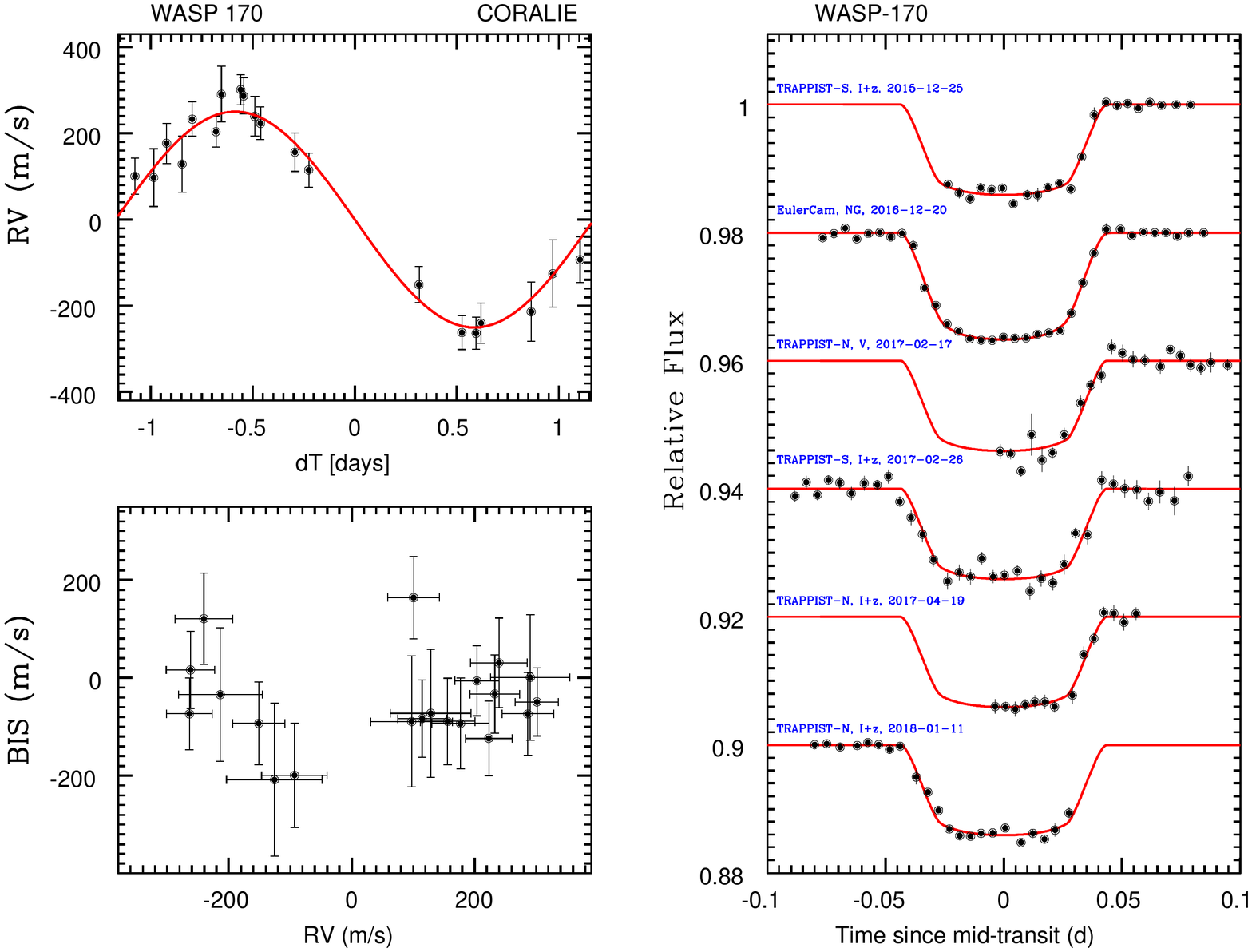}
	\end{center}
	\caption{same as Fig. \ref{wasp161lc} but for WASP-170.}
	\label{wasp170lc}
\end{figure*}

We gathered series of spectra  of the three stars with the CORALIE  spectrograph \citep{Queloz2000} mounted on the 1.2-meter Euler-Swiss  telescope at ESO La Silla Observatory in Chile. An exposure time of 30 min was used for each of these spectroscopic observations. We measured 24 spectra  of WASP-161 between December 2014  and January 2017;  25  spectra of WASP-163 between June 2015  and May 2017; and 20  spectra of WASP-170 between February 2015  and May 2017. We applied the cross correlation technique described in \cite{Baranne1996} on the spectra of each star to  measure 
the radial velocities (RVs) presented in Table \ref{table_wasp161-163-170}.  The resulting RV time-series show clear sinusoidal signals with periods and phases in good agreement with those deduced from the WASP transit detections (Fig. \ref{wasp161lc}, \ref{wasp163lc}, and \ref{wasp170lc}).  

For each star, the bisector spans (BS, \citealt{Queloz2001}) of the cross correlation functions (CCF) have standard deviations close to their average errors (122 $vs$ 80 ms$^{-1}$, 116 $vs$ 125 ms$^{-1}$ and  87 $vs$ 97 ms$^{-1}$ for WASP-161, WASP-163 and  for WASP-170 respectively). Furthermore, a linear regression analysis do not show any significant correlation with between these BS and the corresponding RVs, the computed slopes being $-0.02\pm0.16$, $0.07\pm0.09$, and $0.01\pm0.11$ for, respectively, WASP-161, WASP-163 and WASP-170 (Fig. \ref{wasp161lc}, \ref{wasp163lc}, and \ref{wasp170lc}).  This absence of correlation enables us to discard the scenario of a blended eclipsing binary (BEB). Indeed, if the orbital signal of a BEB was causing a clear periodic wobble of the sum of its CCF(s) and the one of the target, then it should also create a significant periodic distortion of its shape, resulting in variations of the BS in phase with those of the RV, and with the same order of magnitude \cite{Torres2004}. 

\section{Data Analysis} \label{sec:analysis}

\subsection{Spectroscopic analysis} \label{sec:spectro_analysis}

For each host star, we co-added the CORALIE spectra to produce a combined spectrum with an average $S/N$ per pixel between 50 and 100. We analyzed each combined spectrum with the technique described by \cite{Doyle2013} to determine the following stellar atmospheric parameters: the effective temperature $T_{eff}$, the surface gravity $\log g$, the lithium abundance $\log{A}$(Li), the metallicity $[Fe/H]$, and the projected rotational velocity $v\sin i$. $v\sin i$ was constrained using  the calibration of \cite{Doyle2014}, assuming macroturbulence values  of 5.31 km.s$^{-1}$, 3.59 km.s$^{-1}$ and 3.74 km.s$^{-1}$ for WASP-161, WASP-163 and WASP-170 respectively. The results of this spectral analysis are shown in Table \ref{mcmcwasp161163170}.

\subsection{RVs + light curves analysis} \label{sec:mcmc}

We performed a global analysis of the RVs (Table \ref{table_wasp161-163-170}) and transit light curves (Table \ref{allmcmc}) with the MCMC (Markov-Chain Monte Carlo) algorithm described by \cite{Gillon2012} to determinate the parameters of each planetary system. While the CORALIE RVs were modeled with a classical Keplerian model (e.g. \citealt{Murray2010}), the transit light curves were represented by the transit model of \cite{Mandel2002}, assuming a quadratic limb-darkening law, multiplied by a baseline model consisting of a polynomial function of one or several external parameters (time, background, airmass, etc., see Table \ref{allmcmc}).  The selection of the model used for each time-series was based on the minimization of the Bayesian Information Criterium (BIC, \citealt{schwarz1978}).   

TRAPPIST-North and TRAPPIST-South telescopes are equipped with German equatorial mounts that have to rotate of $180^\circ$ at meridian, resulting in different positions of the stars' images on the detector after the flip, translating into an offset of the fluxes in the light curves. For the corresponding light curves, a normalization offset at the time of the flip was thus added to the assumed model (Table \ref{allmcmc}).

For each system, the "jump" parameters of the Markov Chains, i.e. the parameters perturbed at each step of the chains,  were the transit duration, depth, and impact parameter ($W$, $dF$ and $b$, respectively), the orbital period $P$, the mid-transit time $T_0$, the parameters $\sqrt{e} \cos \omega$ and $\sqrt{e} \sin \omega$ (with $\omega$ the argument of periastron and $e$ the orbital eccentricity), the parameter $K2 = K \sqrt{1-e^2} P^{1/3}$ (with $K$ is the RV semi-amplitude), and the stellar metallicity $[Fe/H]$ and  effective temperature $T_{eff}$. In addition, for each filter, the combinations, $c_1 = 2\times u_1+u_2$ and $c_2 =  u_1-2\times u_2$ were also jump parameters, $u_1$ and $u_2$ being the linear and quadratic limb-darkening coefficients. Normal prior probability distribution functions (PDFs) based on the theoretical tables of \cite{Claret2000} were assumed for $u_1$ and $u_2$ (Table \ref{tableLD}). For $T_{eff}$ and $[Fe/H]$, gaussian PDFs based on the values + errors derived from our spectral analysis (Table \ref{mcmcwasp161163170}) were used. For the other jump parameters,  uniform prior PDFs were assumed (e.g. $e \geq 0$, $b \geq 0$). 

Each global analysis was composed of three Markov Chains of $10^5$ steps whose convergence was checked using the statistical test presented by \cite{Gelman1992}.  The correlation of the noise present in the light curves was taken into account by rescaling the errors as described by \cite{Gillon2012}. For the RVs, the quadratic difference between the mean error of the measurements and the standard deviation of the best-fit residuals were computed as 32.6 m.s$^{-1}$, 54.1 m.s$^{-1}$, and 42.8 m.s$^{-1}$ for  WASP-161, WASP-163 and  WASP-170, respectively. These 'jitter' noises were added quadratically to the errors.  

At each step of the Markov chains, a value for the stellar density $\rho_\ast$ was computed from $dF$, $b$, $W$, $P$, $\sqrt{e} \cos \omega$, and $\sqrt{e} \sin \omega$ (see, e.g., \citealt{Winn2010}). This value of $\rho_\ast$ was then used in combination with the values for $T_{eff}$ and $[Fe/H]$ to compute a value for the stellar mass $M_\ast$ from the empirical calibration of \citep{Enoch2010}.  Two MCMC analyses were performed for each system, one assuming an eccentric orbit and one assuming a circular one. The Bayes factors, computed as $\exp(-\Delta BIC/2)$, were largely ($>1000$) in favor of a circular model for the three systems, and we thus adopted the circular solution for all of them. These solutions are presented in Table \ref{mcmcwasp161163170}.  The non-circular solutions enable us to estimate the $3\sigma$ upper limits on the orbital eccentricity as 0.43, 0.13 and 0.23 for respectively, WASP-161 b, WASP-163 b and WASP-170 b. 

As a sanity check of our results, we also estimated the stellar radius $R_\star$ from  the star's parallax determined by Gaia \citep{Gaia2018}, its effective temperature $T_{eff}$, and its bolometric magnitude $M_{bol}$, using the equations:
\begin{equation}
M_v = V_{mag} - 5\log_{10}(d/10),
\end{equation}
\begin{equation}
M_{bol} = M_v + BC,
\end{equation}
\begin{equation}
L_\star/L_\odot = 10^{0.4 (4.74-M_{bol})},
\end{equation}
\begin{equation}
R_\star = L_\star/4\pi\sigma T_{eff}^4, 
\end{equation}

where $M_v$ is the absolute visual magnitude, $BC$ the bolometric correction \citep{Pecaut2013}, $d$  the distance in parsec $pc$, $L_\star$ is the star luminosity, and $\sigma$ the Stefan-Boltzmann constant. We estimated the error on $R_\star$ by propagating the errors of all other parameters. We obtained $1.55 \pm 0.08$ $R_\odot$ for WASP-161, $0.86 \pm 0.07$ $R_\odot$ for WASP-163, and $0.91 \pm 0.06$ $R_\odot$ for WASP-170, in good agreement with our MCMC results shown in table  \ref{mcmcwasp161163170}. 

\blfootnote{Our photometric and radial-velocity data will be available in the web site, \url{http://cdsarc.u-strasbg.fr} }
	
	\section{Stars Rotation periods} \label{sec:rotation}
	
\begin{figure}
		\centering
		\includegraphics[scale=1.1]{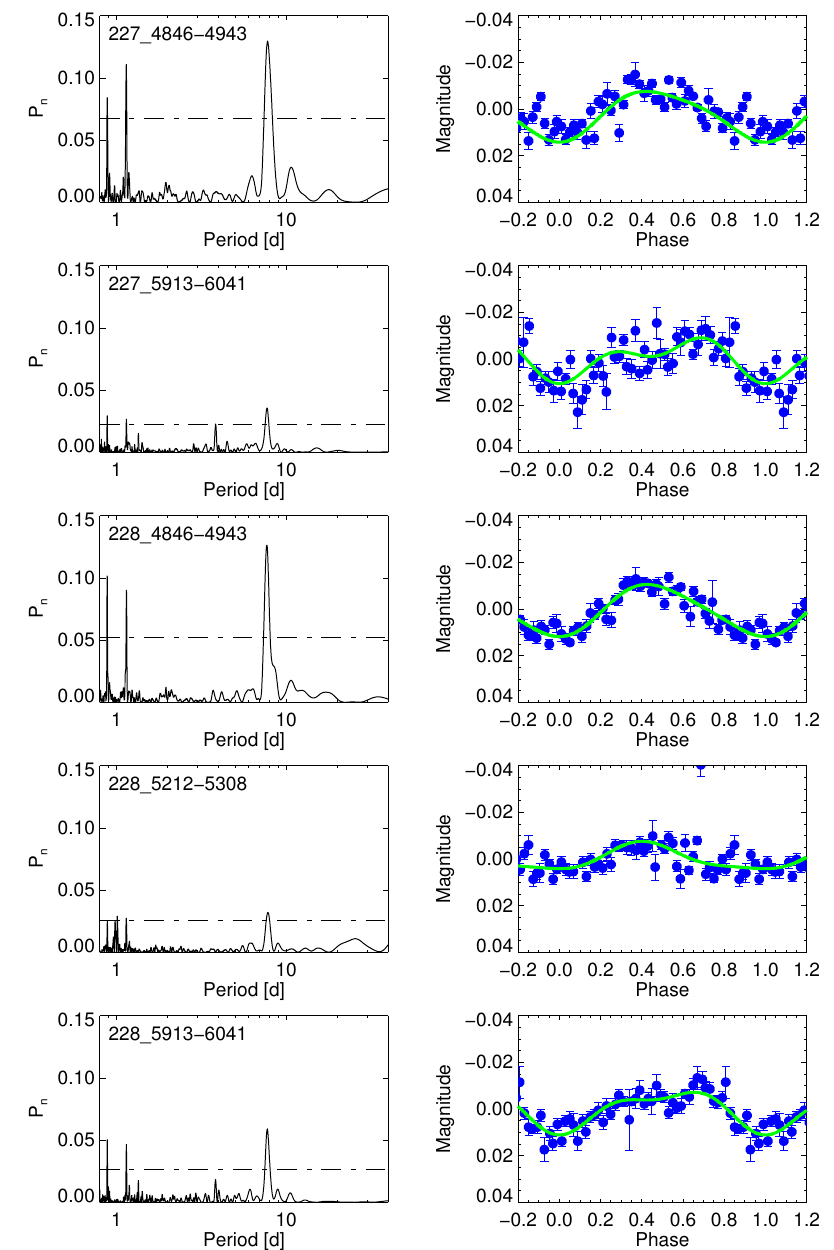}
		\caption{{\it Left panel}: Periodograms of
			WASP-170. Horizontal lines indicate false-alarm probability levels 0.1, 0.01 and 0.001. {\it Right panel}: Light curves binned in the blue points on the assumed rotation period of 7.75 days with second-order harmonic series fit by least squares in the green lines.}
		\label{lcperiodogram}
	\end{figure}

	The WASP-170 light curve from WASP-South show a quasi-periodic modulation with an amplitude of about 0.6 \% and a period of about 7.8 days. We assume this is due to the star spots (i.e. the combination of the star rotation and the magnetic activity). The rotational modulation of each star  was estimated by using the sine-wave fitting method described in \cite{Maxted2011}.  The star variability due to star spots is not expected to be coherent on long time-scales as a consequence of the finite lifetime of the stars spots and differential rotation in the photosphere so we analyzed the WASP-170 data separately. We analyzed  separately the WASP-170 data from each camera used, so that we can estimate the reliability of the results. The transit signal was removed from the data to calculating the periodograms by subtracting a simple transit model from the light curve.  We calculated the periodograms over uniformly spaced frequencies from 0 to 1.5 cycles/day. The False Alarm Probability (FAP) is calculated by the boot-starp Monte Carlo as described in \cite{Maxted2011}. The results are presented in the table \ref{periodogram}, and the periodograms and light curves are shown in the figure \ref{lcperiodogram}. The rotation period value we obtain is  $P_{rot} = 7.75\pm0.02$ days from the clear signal near 7.8 days in 5 out of 7 data sets.   From the stellar radius estimated and rotation period implies the value of $V_{rot} \sin I = 6.1 \pm 0.3$ km s$^{-1}$, assuming that the axises of the star and  the planet orbital  are approximately aligned,  in the good agreement with our spectroscopic analysis ($vsini\ = 5.6 \pm 1.0$ km s$^{-1}$, see the table \ref{mcmcwasp161163170}).
We modeled the rotational modulation in the light curves for each camera and season with the rotation period fixed at $P_{rot} = 7.75$ d using the least-squares fit of a sinusoidal function and its first harmonic.  Similar analysis for WASP-161 and WASP-163.

	\begin{table}[h!]
		\centering
		\begin{tabular}{cccccc}
			\hline
			Camera &  Dates  &  $N$  & $P$ [d]  & $a$ [mmag]  & FAP \\
			 & (JD-2450000) &&&&\\
			\hline \hline
			227  & 4846-4943 & 4899 & 7.780  &  0.010 & $< 10^{-4}$ \\
			227 & 5567-5675 & 2407 & 3.978 & 0.005 & 0.15 \\
			227 & 5913-6041 & 2794 & 7.725 & 0.007 &  $< 10^{-4}$ \\
			228  & 4846-4943 & 5283 & 7.703  &  0.011 & $< 10^{-4}$ \\
			228 & 5212-5308 & 4747 & 7.813 & 0.006 & 0.002 \\
			228 & 5613-5676 & 2651 & 4.073 & 0.003 &  1.00 \\		
			228 & 5913-6041 & 3649 & 7.747 & 0.008 & $< 10^{-4}$  \\
			\hline
		\end{tabular}
		\caption{Periodogram analysis of the WASP light curves for WASP-170.  $N$: the number of observations used in our analysis, $a$: the semi-amplitude of the best-fit sine wave at the period $P$ found in the periodogram with false-alarm probability FAP.}
		\label{periodogram}
	\end{table}
	
	\begin{table*}[h!]
		{\renewcommand{\arraystretch}{1.4}
			\small
			\begin{tabular}{lccccccccccc}
				\hline
				Target  &  Night &  Telescope & Filter  & $N_p$  & $T_{exp}$ (s) & Baseline function & $\sigma$ (\%) & $\sigma_{7.2m}$(\%) & $\beta_w$ & $\beta_r$ & $CF$ \\
				\hline\hline
				WASP-161 & 2016-01-28& TRAPPIST-S & Sloan-$z'$  & 938 & 10 & $p(t+xy+o)$          & 0.37 & 0.051 & 1.29 & 1.08 & 1.39\\
				WASP-161 & 2017-12-20& TRAPPIST-N & Sloan-$z'$  & 902 & 10 & $p(t+b)$               & 0.43 & 0.072 & 1.14 & 1.05 & 1.20\\
				WASP-161 & 2018-01-05& SPECULOOS & Sloan-$z'$  & 1235 & 10 & $p(xy)$                 & 0.44 & 0.087 & 1.22 & 1.44 & 1.72\\
				WASP-161 & 2018-02-12& TRAPPIST-N & Sloan-$z'$  & 892 & 10 & $p(a)$                     & 0.46 & 0.054 & 1.14 & 1.20 & 1.37 \\
				&&&&&&&&&&&\\
				WASP-163 & 2014-09-06 & TRAPPIST-S & $I+z$  & 345 & 12 & $p(t) $ & 0.33 & 0.006 & 1.06 & 1.00& 1.06 \\
				WASP-163 & 2016-06-27 & NITES & Johnson-R  & 443  & 30   & $p(t)$                   & 0.41 & 0.012 & 1.71 & 1.00 & 1.71\\
				WASP-163 & 2016-06-27 & EulerCam & RG &  170 & 60  & $p(t+f+b)$          & 0.11 & 0.005 & 1.20 & 1.10 & 1.31\\
				WASP-163 & 2016-07-05 & TRAPPIST-S & $I+z$  & 602  & 12 & $p(a+xy)$  & 0.35 & 0.011 & 1.16 & 1.35 & 1.56\\
				WASP-163 & 2016-07-10 & NITES & Johnson-I  & 388 & 30 &  $p(t)$                        & 0.58 & 0.012 & 1.55 & 1.18 & 1.83\\
				WASP-163 & 2017-04-24 & TRAPPIST-N & $I+z$  & 487 & 12 & $p(t+xy+o)$  & 0.31 & 0.008 & 1.02 & 1.07 & 1.09\\
				WASP-163 & 2017-05-02 & TRAPPIST-N & $I+z$  & 213 & 12 & $p(b)$                  & 0.54 & 0.009 & 0.90 & 1.00 & 0.90\\
				WASP-163 & 2017-06-13 & TRAPPIST-N & $I+z$  & 557 & 14 & $p(t+f)$           & 0.69 & 0.021 & 0.87 & 1.16 & 1.01\\
				&&&&&&&&&&&\\
				WASP-170 & 2015-12-25& TRAPPIST-S & $I+z$  & 359 & 15 & $p(f)$                    & 0.29 & 0.008 & 1.04 & 1.05 & 1.09\\
				WASP-170 & 2016-12-20& EulerCam & NGTS & 207 & 40 & $p(t)$                         & 0.11 & 0.005 & 1.49 & 1.13 & 1.68\\
				WASP-170 & 2017-02-17& TRAPPIST-N & Johnson-V & 239 & 20 & $p(t)$   & 0.46 & 0.013 & 1.22 & 1.00 & 1.22\\
				WASP-170 & 2017-02-26& TRAPPIST-S & $I+z$ & 545  & 15 & $p(t+a)$    & 0.51 & 0.015 & 1.32 & 1.16 & 1.53\\
				WASP-170 & 2017-04-19 & TRAPPIST-N & $I+z$ & 186  & 15 & $p(a+xy)$  & 0.41 & 0.008 & 0.99 & 1.00 & 0.99\\
				WASP-170 & 2018-01-11 & TRAPPIST-N  & $I+z$ & 315  & 15 & $p(t)$   & 0.26 & 0.007 & 0.75 & 1.11 & 0.83 \\
				\hline
		\end{tabular}}
		\caption{The table shows for each light curve the date ,  telescope, filter,  number of data points, the exposure time, the selected baseline function, the RMS of the best-fit residuals, the deduced values for $\beta_w$, $\beta_r$ and $CF =  \beta_w \times \beta_r$. For the baseline function, $p(\epsilon^N)$, denotes, respectively, a $N-$order polynomial function of time ($\epsilon = t$), airmass ($\epsilon = a$), full-width at half maximum ($\epsilon=f$), background ($\epsilon = b$), and $x$ and $y$ positions ($\epsilon = xy$). The symbol $o$ demotes an offset fixed at the time of the meridian flip.}
		\label{allmcmc}
	\end{table*}

	\section{Stellar evolution modeling} \label{sec:stellar}
	
	We estimated the mass and age of the host stars using the software {\it BAGEMASS}\footnote{http://sourceforge.net/projects/bagemass} based on the Bayesian method described in \cite{Maxted2015}. The models used in the software {\it BAGEMASS} were calculated using the   {\it GARSTEC} stellar evolution code as described in \cite{Weiss2008}. The deduced stellar masses and ages  calculated are shown in the table \ref{AgeMass}. The inferred masses are in good agreement with the ones resulting from our global MCMC analysis (see Table \ref{mcmcwasp161163170}).

	\begin{table}[h!]
		\centering
		\caption{Stellar mass and age estimates from the software {\it BAGEMASS}.}
		{\renewcommand{\arraystretch}{1.5}
			\resizebox{0.5\textwidth}{!}{
				\begin{tabular}{ccc}
					\hline
					Star &  Mass  [$M_\odot$]  & Age [Gyr] \\
					\hline \hline
					WASP-161 A & $1.42\pm 0.05 $ (1.40) & $2.4\pm 0.4$ (2.4)\\
					WASP-163 A & $0.87 \pm 0.06$ (0.78)  & $11.4 \pm 3.5$ (17.4)$^\dag$ \\
					WASP-170 A & $0.99 \pm 0.07$ (1.03) & $4.8\pm 3.1$ (2.9)\\
					\hline
					$^\dag$ Best fit occurs at edge of model grid.
		\end{tabular}}}
		\label{AgeMass}
	\end{table}
	
	\begin{table}[h!]
			\begin{center}		 
		{\renewcommand{\arraystretch}{1.2}
			\resizebox{0.49\textwidth}{!}{
				\begin{tabular}{cccc}
					\hline
					LD coefficient         &   WASP-161 &          WASP-163            &        WASP-170  \\
					\hline\hline
					$u_{1, z'}$     &  $0.184 \pm  0.011$    & - & - \\
					$u_{2, z'}$     &   $0.300 \pm 0.005$ & - &  - \\
					$u_{1, I+z}$              & - &   $0.207\pm 0.012$     &   $0.2539\pm 0.0202$  \\
					$u_{2, I+z}$              & - &  $0.297\pm 0.010$     & $ 0.2788\pm 0.0152 $  \\
					$u_{1, Johnson-I}$    & - &  $ 0.331 \pm 0.034 $  &  $0.2727\pm 0.0321$ \\
					$u_{2, Johnson-I}$   & - & $0.251 \pm 0.019$     & $0.2805 \pm 0.0158$ \\
					$u_{1, Johnson-R}$   & - & $ 0.420\pm 0.043 $  &  -  \\
					$u_{2, Johnson-R}$  & - &  $0.248\pm 0.027$    & - \\
					$u_{1, Johnson-V}$   & -  &           -                       &  $0.437 \pm 0.044$ \\
					$u_{2, Johnson-V}$   & -  &           -                        &  $0.271 \pm 0.025$ \\
					\hline
		\end{tabular}}}
		\caption{The quadratic limb-darkening  (LD) coefficients $u_1$ and $u_2$ used in our MCMC analysis. }
		\label{tableLD}
		\end{center}
	\end{table}

	\section{Discussion} \label{sec:conclusion}
	
\begin{figure}
		\includegraphics[scale=0.35]{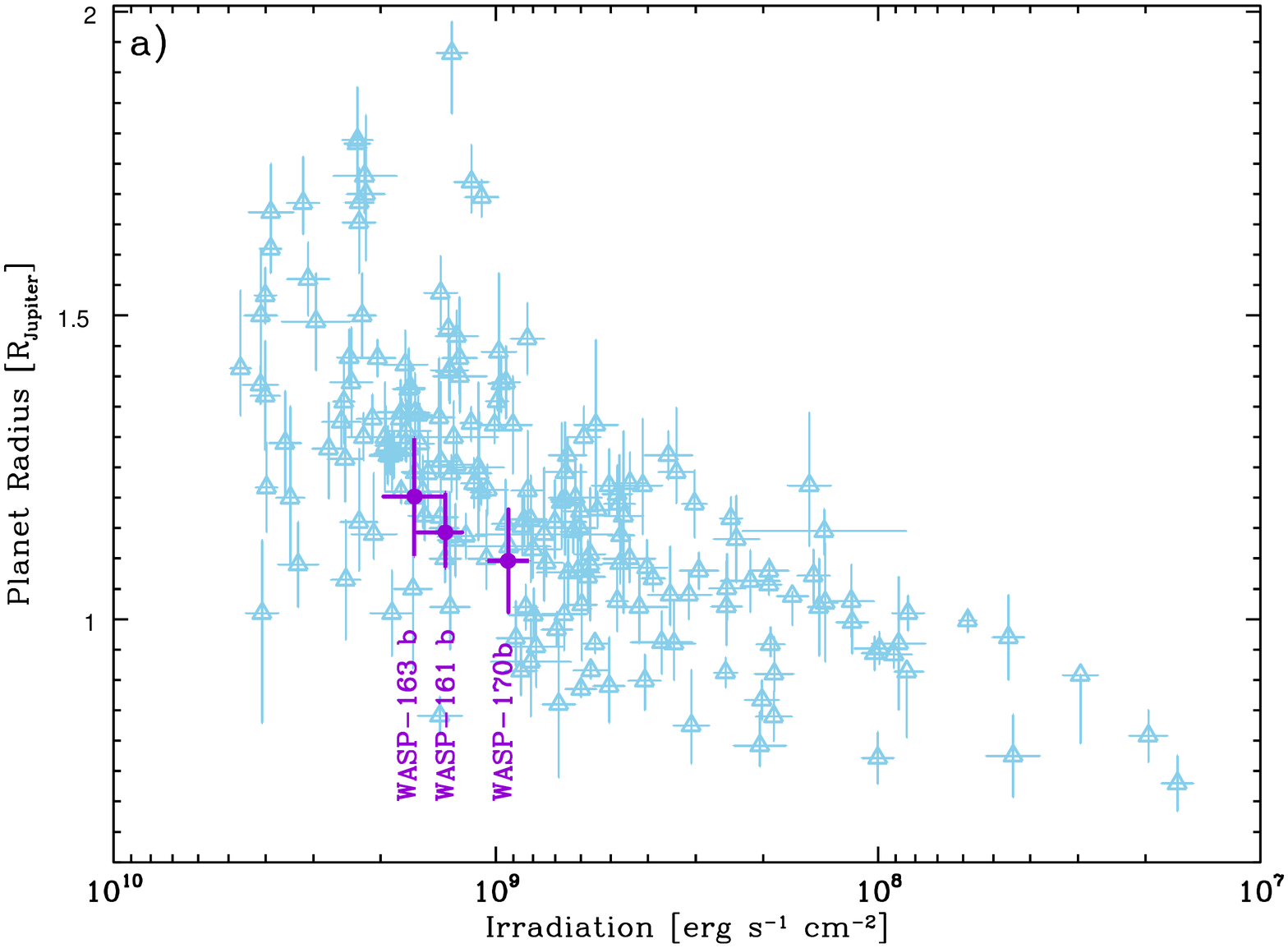}   \includegraphics[scale=0.35]{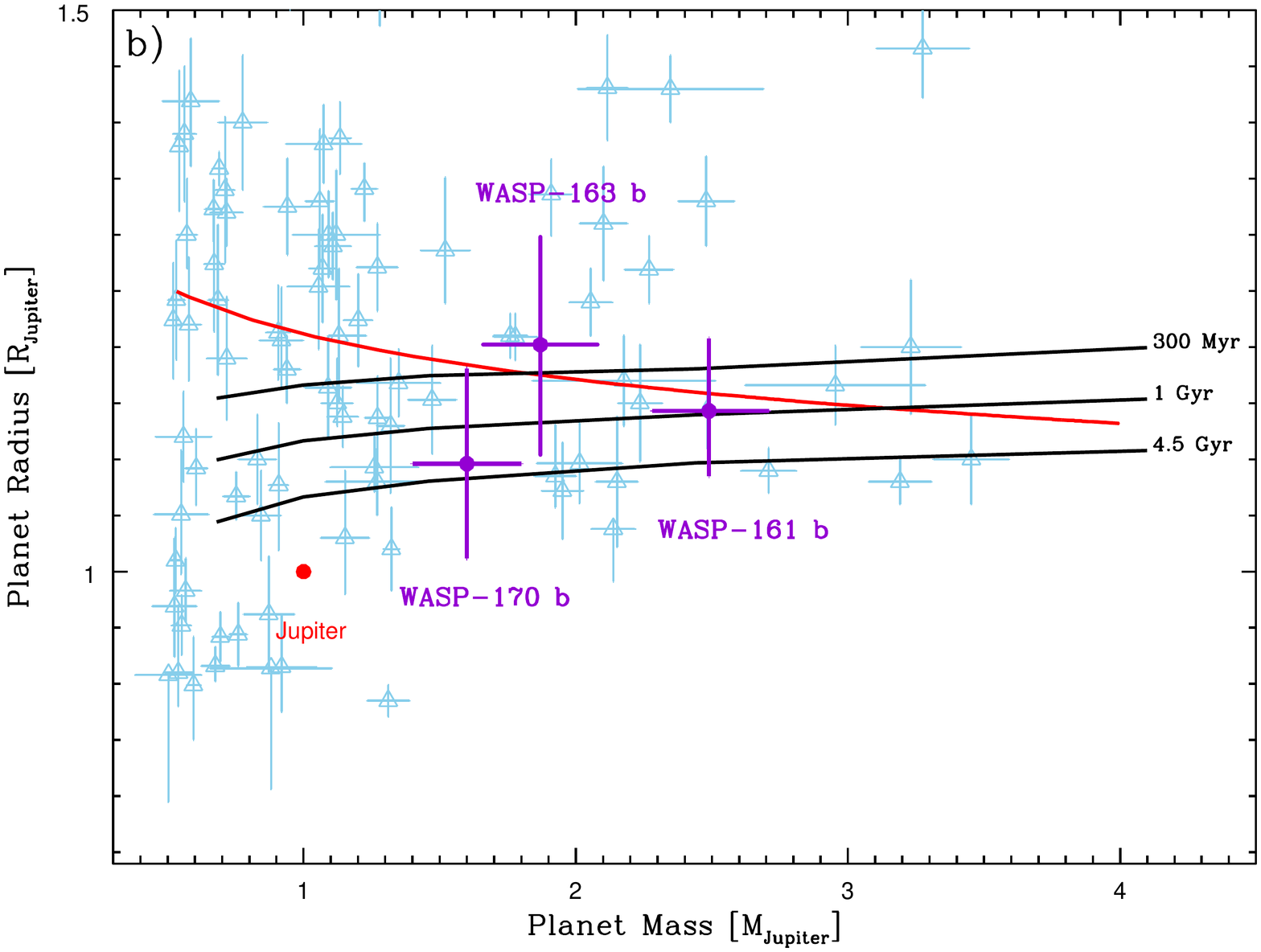}
		\caption{\textbf{a)}: Irradiation-radius diagram and \textbf{b)}: mass-radius diagram for the known transiting planets with masses ranging from $0.5$ to $4$ $M_{Jupiter}$  (data from \href{http://exoplanets.org/}{exoplanets.org} are shown as skyblue triangles with error bars). The planets WASP-161 b, WASP-163 b and WASP-170 b are shown in violet. In \textbf{b)} the black lines present models of irradiated giant planets with semi-major axes of 0.045 AU, core masses of 25 $M_\oplus$, and ages  of  300 Myrs, 1 Gyrs and 4.5 Gyrs \citep{Fortney2007}. The empirical law of \cite{Weiss2013} is also plotted as a red line.}
		\label{RpMpIrrad}
	\end{figure}

	WASP-161\ b, WASP-163\ b and WASP-170\ b are  planets slightly larger ($ 1.14\pm0.06$ $R_{Jup}$, $1.2\pm0.1$ $R_{Jup}$, and $1.10\pm0.09$ $R_{Jup}$) and more massive ($2.5\pm0.2$ $M_{Jup}$, $1.9\pm0.2$ $M_{Jup}$, and $1.6\pm0.2$ $M_{Jup}$) than Jupiter.  Given their masses and their large irradiations (Fig. \ref{RpMpIrrad} \textbf{a}), their radii are well reproduced by the models of \cite{Fortney2007}, assuming a core mass of a few dozens of $M_\oplus$ and ages larger than a few hundreds Myr  (Fig. \ref{RpMpIrrad} \textbf{b}). 
	
	The empirical relationship derived by  \cite{Weiss2013} for  planets more massive than $150 M_\oplus$, $ R_p/R_\oplus = 2.45 (M_p/M_\oplus)^{-0.039\pm 0.01} (F/\text{erg } \text{s}^{-1} \text{cm}^{-2})^{0.094}$ predicts radii of  $1.16 \pm 0.30$ $R_{Jup}$, $1.20\pm 0.34$ $R_{Jup}$ and $1.15\pm 0.31$ $R_{Jup}$ for WASP-161\,b, 163\,b, and 170\,b, respectively, which are consistent with our measured radii. The three new planets whose discovery is described appear thus to be  'standard' hot Jupiters that do not present a 'radius anomaly' challenging standard models of irradiated gas giants.
	
	The discovery of WASP-161 b, WASP-163 b, and WASP-170 b establishes the new robotic telescope TRAPPIST-North as a powerful Northern facility for the photometric follow-up of transiting exoplanet candidates found by ground-based wide-field surveys like WASP, and soon by the space-based mission TESS \citep{Richer2016TESS}. 
	
	\section{Acknowledgement}
	
	WASP-South is hosted by the SAAO  and we are grateful for their ongoing support
	and assistance. Funding for WASP comes from consortium
	universities and from the UK's Science and Technology Facilities Council. The Euler-Swiss telescope is supported by the Swiss National Science Foundation. TRAPPIST-South	is funded by the Belgian Fund for Scientic Research (FNRS) under the grant FRFC 2.5.594.09.F, with the participation of the Swiss
	National Science Foundation (SNF). MG is FNRS Research Associate, and EJ is FNRS Senior Research Associate. LD acknowledges support from the Gruber Foundation Fellowship. The research leading to these results has	received funding from the European Research Council under the FP/2007-2013 ERC Grant Agreement 336480,	and from the ARC grant for Concerted Research Actions, financed by the Wallonia-Brussels Federation. This work was
	also partially supported by a grant from the Simons Foundation (ID 327127 to Didier Queloz), a grant from the Erasmus+ International Credit Mobility programme (K Barkaoui), as well as by the MERAC foundation (PI Triaud).

	\begin{table*}[h!]
		\begin{tabular}{lcccc||lcccc}
			\hline
			Target    & HID - & RV & $\sigma_{RV}$ & BS   & Target    & HID - & RV & $\sigma_{RV}$ & BS  \\
			&                  2,450,000         & (km s$^{-1}$)      &  (km s$^{-1}$)  & (km s$^{-1}$)  & &    2,450,000   & (km s$^{-1}$)      &  (km s$^{-1}$)  & (km s$^{-1}$) \\    
			\hline\hline
			WASP-163 & 7193.741864	  &   -37.28368		  &   0.11764		  &   	-0.05553 &		WASP-161 &  6995.779435  &	37.47673   &	0.03954	    &    	 0.10140   \\	 
			WASP-163 & 7194.544082	  &   	-37.93075	  &   	0.07383		  &   	0.14349 & 		WASP-161 &  7404.735111  &	37.93140   &	0.04051  	  &       -0.15260 \\	
			WASP-163 & 7221.642393	  &   	-37.90136	  &   	0.07320		  &   	-0.10196 &			WASP-161 &  7421.604076  &	37.83967   &	0.03886 	   &   -0.17254	    \\
			WASP-163 & 7264.560197		  &   -37.21532	  &   	0.08258		  &   	0.30132	 & 		WASP-161 &  7422.600890  &	37.60262   &	0.03490	    &   	 -0.17929		 \\
			WASP-163 & 7265.528948	  &   	-37.85780	  &   	0.08717		  &   	0.20131	& 		WASP-161 &  7423.668930  &	37.52064   &	0.03555	    &     	 -0.28021 \\	
			WASP-163 & 7268.579849		  &   -37.96344		  &   0.07401	&   	0.09958 &	WASP-161 &  7425.646222  &	37.72356   &	0.03456 	   &   -0.04948		 \\			 
			WASP-163 & 7276.497340		  &   -37.95927	  &   	0.10641		  &   	-0.01258 &		WASP-161 &  7426.579130  &	37.96837   &	0.03946	    &    	 -0.16563	 \\	
			WASP-163 & 7277.497313		  &   -37.30844	  &   	0.07177		  &   	-0.11522 &		WASP-161 &  7428.627488  &	37.41733   &	0.03252	    &    	 0.04443		 \\	
			WASP-163 & 7292.517764	  &   	-37.82511	  &   	0.09442		  &   	0.24068 &		WASP-161 &  7451.572650  &	37.47357   &	0.03171	    &    	 -0.18188	 \\		
			WASP-163 & 7293.464761	  &   	-37.14368	  &   	0.07405	    &   	0.24267 &		WASP-161 &  7452.635935  &	37.57245   &	0.03459	    &     	 0.02944	 \\	
			WASP-163 & 7294.524814	  &   	-37.73113	  &   	0.08910		  &   	0.15248 & 		WASP-161 &  7453.567361  &	37.93838   &	0.03395	    &     	 -0.14701 \\	
			WASP-163 & 7484.863204	  &   	-37.46746	  &   	0.03879		&   	-0.03903 &	 WASP-161 &  7457.538982  &	37.46129   &	0.03590	    &     	 0.10144	 \\			 
			WASP-163 & 7486.827998	  &   	-37.42444	  &   	0.04741	      &   	0.16399 &  WASP-161 &  7481.604825  &	37.65253   &	0.04218	    &     	 -0.16457 \\
			WASP-163 & 7487.796538	  &   	-37.79839	  &   	0.04198		  &   	0.09038 & 	WASP-161 &  7485.611602  &	37.80633   &	0.03723	 & 0.06739	 \\
			WASP-163 & 7488.843232	&   	-37.86941	  &   	0.03227  &  0.08838 &WASP-161 &  7669.874869  &	37.96450   &	0.03643	  &  -0.26075 \\	
			WASP-163 & 7523.858463	  &   	-37.41079	  &   	0.04884	&  0.02768 & WASP-161 &  7670.873488  &	37.54791   &	0.03644	 &  0.07926	 \\		
			WASP-163 & 7567.695488	  &   	-37.80440	  &   	0.03823	& 0.05318  & WASP-161 &  7674.868344  &	37.89486   &	0.04877	& 	-0.13512 \\	
			WASP-163 & 7569.757824	  &   	-37.85478	  &   	0.05104 & 0.07520	& WASP-161 &  7716.752403  &	37.48642   &	0.06487	&  -0.04681 \\		
			WASP-163 & 7575.726750		  &   -37.81565	  &   	0.04977	&  -0.13093 & WASP-161 &  7717.747923  &	37.75662   &	0.04854	 &     -0.01920 \\		
			WASP-163 & 7576.662344	  &   	-37.38270	  &   	0.04033	&  0.02235 & WASP-161 &  7718.788841  &	37.80035   &	0.03818	 &  -0.07805 \\	
			WASP-163 & 7577.523403		  &   -37.95890	  &   	0.04961	& 0.01980  & WASP-161 &  7726.778064  &	37.39224   &	0.03876	&  0.01718	 \\ 
			WASP-163 & 7593.681424	  &   	-37.95775	  &   	0.05024	& 0.00629	& WASP-161 &  7746.846331  &	37.51936   &	0.03845	 & -0.04507 \\ 
			WASP-163 & 7652.535369	  &   	-37.33184	  &   	0.03470 &  -0.02346  &  WASP-161 &  7751.735303  &	37.69111   &	0.05137	 &  -0.11653 \\
			WASP-163 & 7823.841592		  &   -38.04752	  &   	0.05155	&  -0.10981  &  WASP-161 &  7761.696319  &	37.93171   &	0.04203	 & -0.35162 \\
			WASP-163 & 7894.742995	  &   	-37.97268	  &   	0.04302 & 0.07214	&  & & & & \\
			\hline \hline
			WASP-170 & 7066.749515  &	30.67098   &		0.04671	 &		0.16402	&   WASP-170 &  7753.676136	 &	30.70862 &		0.03924	 &		-0.12366 \\
			WASP-170 &  7686.843189	 &	31.25138 &		0.06427	 &		-0.08954		&  WASP-170 &  7754.698984 &		31.19850 &		0.04009	 &	-0.08925	\\
			WASP-170 &  7694.846421	 &	30.81621 &		0.04224	 &		-0.09353 &  WASP-170 &  7759.695126	 &	31.20223 &		0.04579	 &		-0.08356	\\
			WASP-170 &  7719.778573	 &	31.25569 &		0.04232	 &		-0.07239  &  WASP-170 &  7760.780900	 &	30.70044 &		0.03683	 &		-0.09317	\\
			WASP-170 &  7721.746763	 &	31.13284 &		0.04660	 &		-0.03291  &  WASP-170 &  7773.792434	 &	31.17400 &		0.03803	 &		0.01621	\\
			WASP-170 &  7723.773046	 &	30.84403 &		0.05319	 &		-0.00621	&   WASP-170 &  7801.545185	 &	31.08175 &		0.06548	 &		-0.07362	\\
			WASP-170 &  7724.787756	 &	31.05367 &		0.03945	 &		0.00095	&    WASP-170 &  7812.635803	 &	30.67711 &		0.06835	 &		0.12073	\\
			WASP-170 &  7726.799025 &		31.25678 &		0.03495	 &		-0.04931  &  WASP-170 &  7825.546120	 &	31.14228 &		0.04425	 &		-0.03448\\
			WASP-170 &  7747.780705	 &	31.16081 &		0.03596	 &		-0.07385	&  WASP-170 &  7859.637199	 &	30.88820 &		0.07800	 &		-0.20830  \\
			WASP-170 &  7749.729288	 &	31.01646 &		0.04208	 &		0.03075	&  WASP-170 &  7883.473343	 &	31.07842 &		0.06692	 &		-0.19927	\\
			\hline
		\end{tabular}
		\caption{CORALIE radial-velocity measurements for WASP-161, WASP-163 and WASP-170 (BS = bisector spans).}
		\label{table_wasp161-163-170}
	\end{table*}

\bibliographystyle{achemso}
\bibliography{biblio}

\providecommand{\latin}[1]{#1}
\providecommand*\mcitethebibliography{\thebibliography}
\csname @ifundefined\endcsname{endmcitethebibliography}
  {\let\endmcitethebibliography\endthebibliography}{}
\begin{mcitethebibliography}{54}
\providecommand*\natexlab[1]{#1}
\providecommand*\mciteSetBstSublistMode[1]{}
\providecommand*\mciteSetBstMaxWidthForm[2]{}
\providecommand*\mciteBstWouldAddEndPuncttrue
  {\def\EndOfBibitem{\unskip.}}
\providecommand*\mciteBstWouldAddEndPunctfalse
  {\let\EndOfBibitem\relax}
\providecommand*\mciteSetBstMidEndSepPunct[3]{}
\providecommand*\mciteSetBstSublistLabelBeginEnd[3]{}
\providecommand*\EndOfBibitem{}
\mciteSetBstSublistMode{f}
\mciteSetBstMaxWidthForm{subitem}{(\alph{mcitesubitemcount})}
\mciteSetBstSublistLabelBeginEnd
  {\mcitemaxwidthsubitemform\space}
  {\relax}
  {\relax}

\bibitem[{Mayor} and {Queloz}(1995){Mayor}, and {Queloz}]{mayor1995}
{Mayor},~M.; {Queloz},~D. \emph{\nat} \textbf{1995}, \emph{378}, 355--359\relax
\mciteBstWouldAddEndPuncttrue
\mciteSetBstMidEndSepPunct{\mcitedefaultmidpunct}
{\mcitedefaultendpunct}{\mcitedefaultseppunct}\relax
\EndOfBibitem
\bibitem[{Charbonneau} \latin{et~al.}(2000){Charbonneau}, {Brown}, {Latham},
  and {Mayor}]{charbonneau2000}
{Charbonneau},~D.; {Brown},~T.~M.; {Latham},~D.~W.; {Mayor},~M. \emph{\apjl}
  \textbf{2000}, \emph{529}, L45--L48\relax
\mciteBstWouldAddEndPuncttrue
\mciteSetBstMidEndSepPunct{\mcitedefaultmidpunct}
{\mcitedefaultendpunct}{\mcitedefaultseppunct}\relax
\EndOfBibitem
\bibitem[{Henry} \latin{et~al.}(2000){Henry}, {Marcy}, {Butler}, and
  {Vogt}]{Henry2000}
{Henry},~G.~W.; {Marcy},~G.~W.; {Butler},~R.~P.; {Vogt},~S.~S. \emph{\apjl}
  \textbf{2000}, \emph{529}, L41--L44\relax
\mciteBstWouldAddEndPuncttrue
\mciteSetBstMidEndSepPunct{\mcitedefaultmidpunct}
{\mcitedefaultendpunct}{\mcitedefaultseppunct}\relax
\EndOfBibitem
\bibitem[{Winn} and {Fabrycky}(2015){Winn}, and {Fabrycky}]{Winn2015}
{Winn},~J.~N.; {Fabrycky},~D.~C. \emph{\araa} \textbf{2015}, \emph{53},
  409--447\relax
\mciteBstWouldAddEndPuncttrue
\mciteSetBstMidEndSepPunct{\mcitedefaultmidpunct}
{\mcitedefaultendpunct}{\mcitedefaultseppunct}\relax
\EndOfBibitem
\bibitem[{Fortney} \latin{et~al.}(2007){Fortney}, {Marley}, and
  {Barnes}]{Fortney2007}
{Fortney},~J.~J.; {Marley},~M.~S.; {Barnes},~J.~W. \emph{\apj} \textbf{2007},
  \emph{659}, 1661--1672\relax
\mciteBstWouldAddEndPuncttrue
\mciteSetBstMidEndSepPunct{\mcitedefaultmidpunct}
{\mcitedefaultendpunct}{\mcitedefaultseppunct}\relax
\EndOfBibitem
\bibitem[{Correia} and {Laskar}(2010){Correia}, and {Laskar}]{Correia2010}
{Correia},~A.~C.~M.; {Laskar},~J. \emph{\icarus} \textbf{2010}, \emph{205},
  338--355\relax
\mciteBstWouldAddEndPuncttrue
\mciteSetBstMidEndSepPunct{\mcitedefaultmidpunct}
{\mcitedefaultendpunct}{\mcitedefaultseppunct}\relax
\EndOfBibitem
\bibitem[{Chang} \latin{et~al.}(2010){Chang}, {Liu}, {Tang}, {Chen}, {Shao},
  and {Huang}]{Chang2010}
{Chang},~C.; {Liu},~G.~Z.; {Tang},~C.~X.; {Chen},~C.~H.; {Shao},~H.;
  {Huang},~W.~H. \emph{Applied Physics Letters} \textbf{2010}, \emph{96},
  111502\relax
\mciteBstWouldAddEndPuncttrue
\mciteSetBstMidEndSepPunct{\mcitedefaultmidpunct}
{\mcitedefaultendpunct}{\mcitedefaultseppunct}\relax
\EndOfBibitem
\bibitem[{Winn}(2010)]{Winn2010}
{Winn},~J.~N. \emph{ArXiv e-prints} \textbf{2010}, \relax
\mciteBstWouldAddEndPunctfalse
\mciteSetBstMidEndSepPunct{\mcitedefaultmidpunct}
{}{\mcitedefaultseppunct}\relax
\EndOfBibitem
\bibitem[{Deming} and {Seager}(2009){Deming}, and {Seager}]{Deming2009}
{Deming},~D.; {Seager},~S. \emph{\nat} \textbf{2009}, \emph{462},
  301--306\relax
\mciteBstWouldAddEndPuncttrue
\mciteSetBstMidEndSepPunct{\mcitedefaultmidpunct}
{\mcitedefaultendpunct}{\mcitedefaultseppunct}\relax
\EndOfBibitem
\bibitem[{Seager} and {Deming}(2010){Seager}, and {Deming}]{Seager2010}
{Seager},~S.; {Deming},~D. \emph{\araa} \textbf{2010}, \emph{48},
  631--672\relax
\mciteBstWouldAddEndPuncttrue
\mciteSetBstMidEndSepPunct{\mcitedefaultmidpunct}
{\mcitedefaultendpunct}{\mcitedefaultseppunct}\relax
\EndOfBibitem
\bibitem[{Sing} \latin{et~al.}(2016){Sing}, {Fortney}, {Nikolov}, {Wakeford},
  {Kataria}, {Evans}, {Aigrain}, {Ballester}, {Burrows}, {Deming},
  {D{\'e}sert}, {Gibson}, {Henry}, {Huitson}, {Knutson}, {Lecavelier Des
  Etangs}, {Pont}, {Showman}, {Vidal-Madjar}, {Williamson}, and
  {Wilson}]{Sing2016}
{Sing},~D.~K. \latin{et~al.}  \emph{\nat} \textbf{2016}, \emph{529},
  59--62\relax
\mciteBstWouldAddEndPuncttrue
\mciteSetBstMidEndSepPunct{\mcitedefaultmidpunct}
{\mcitedefaultendpunct}{\mcitedefaultseppunct}\relax
\EndOfBibitem
\bibitem[{Crossfield}(2015)]{Crossfield2015}
{Crossfield},~I.~J.~M. \emph{\pasp} \textbf{2015}, \emph{127}, 941\relax
\mciteBstWouldAddEndPuncttrue
\mciteSetBstMidEndSepPunct{\mcitedefaultmidpunct}
{\mcitedefaultendpunct}{\mcitedefaultseppunct}\relax
\EndOfBibitem
\bibitem[Pollacco \latin{et~al.}(2006)Pollacco, Skillen, Cameron, Christian,
  Hellier, Irwin, Lister, Street, West, Anderson, Clarkson, Deeg, Enoch, Evans,
  Fitzsimmons, Haswell, Hodgkin, Horne, Kane, Keenan, Maxted, Norton, Osborne,
  Parley, Ryans, Smalley, Wheatley, and Wilson]{Pollacco2006}
Pollacco,~D.~L. \latin{et~al.}  \emph{Publications of the Astronomical Society
  of the Pacific} \textbf{2006}, \emph{118}, 1407--1418\relax
\mciteBstWouldAddEndPuncttrue
\mciteSetBstMidEndSepPunct{\mcitedefaultmidpunct}
{\mcitedefaultendpunct}{\mcitedefaultseppunct}\relax
\EndOfBibitem
\bibitem[{Collier Cameron} \latin{et~al.}(2007){Collier Cameron}, {Wilson},
  {West}, {Hebb}, {Wang}, {Aigrain}, {Bouchy}, {Christian}, {Clarkson},
  {Enoch}, {Esposito}, {Guenther}, {Haswell}, {H{\'e}brard}, {Hellier},
  {Horne}, {Irwin}, {Kane}, {Loeillet}, {Lister}, {Maxted}, {Mayor}, {Moutou},
  {Parley}, {Pollacco}, {Pont}, {Queloz}, {Ryans}, {Skillen}, {Street}, {Udry},
  and {Wheatley}]{Cameron2007}
{Collier Cameron},~A. \latin{et~al.}  \emph{\mnras} \textbf{2007}, \emph{380},
  1230--1244\relax
\mciteBstWouldAddEndPuncttrue
\mciteSetBstMidEndSepPunct{\mcitedefaultmidpunct}
{\mcitedefaultendpunct}{\mcitedefaultseppunct}\relax
\EndOfBibitem
\bibitem[{Hellier} \latin{et~al.}(2011){Hellier}, {Anderson},
  {Collier-Cameron}, {Miller}, {Queloz}, {Smalley}, {Southworth}, and
  {Triaud}]{Hellier2011}
{Hellier},~C.; {Anderson},~D.~R.; {Collier-Cameron},~A.; {Miller},~G.~R.~M.;
  {Queloz},~D.; {Smalley},~B.; {Southworth},~J.; {Triaud},~A.~H.~M.~J.
  \emph{\apjl} \textbf{2011}, \emph{730}, L31\relax
\mciteBstWouldAddEndPuncttrue
\mciteSetBstMidEndSepPunct{\mcitedefaultmidpunct}
{\mcitedefaultendpunct}{\mcitedefaultseppunct}\relax
\EndOfBibitem
\bibitem[{Hellier} \latin{et~al.}(2012){Hellier}, {Anderson}, {Collier
  Cameron}, {Doyle}, {Fumel}, {Gillon}, {Jehin}, {Lendl}, {Maxted}, {Pepe},
  {Pollacco}, {Queloz}, {S{\'e}gransan}, {Smalley}, {Smith}, {Southworth},
  {Triaud}, {Udry}, and {West}]{Hellier2012}
{Hellier},~C. \latin{et~al.}  \emph{\mnras} \textbf{2012}, \emph{426},
  739--750\relax
\mciteBstWouldAddEndPuncttrue
\mciteSetBstMidEndSepPunct{\mcitedefaultmidpunct}
{\mcitedefaultendpunct}{\mcitedefaultseppunct}\relax
\EndOfBibitem
\bibitem[{Collier Cameron} \latin{et~al.}(2006){Collier Cameron}, {Pollacco},
  {Street}, {Lister}, {West}, {Wilson}, {Pont}, {Christian}, {Clarkson},
  {Enoch}, {Evans}, {Fitzsimmons}, {Haswell}, {Hellier}, {Hodgkin}, {Horne},
  {Irwin}, {Kane}, {Keenan}, {Norton}, {Parley}, {Osborne}, {Ryans}, {Skillen},
  and {Wheatley}]{Cameron2006}
{Collier Cameron},~A. \latin{et~al.}  \emph{\mnras} \textbf{2006}, \emph{373},
  799--810\relax
\mciteBstWouldAddEndPuncttrue
\mciteSetBstMidEndSepPunct{\mcitedefaultmidpunct}
{\mcitedefaultendpunct}{\mcitedefaultseppunct}\relax
\EndOfBibitem
\bibitem[{Collier Cameron} \latin{et~al.}(2007){Collier Cameron}, {Wilson},
  {West}, {Hebb}, {Wang}, {Aigrain}, {Bouchy}, {Christian}, {Clarkson},
  {Enoch}, {Esposito}, {Guenther}, {Haswell}, {H{\'e}brard}, {Hellier},
  {Horne}, {Irwin}, {Kane}, {Loeillet}, {Lister}, {Maxted}, {Mayor}, {Moutou},
  {Parley}, {Pollacco}, {Pont}, {Queloz}, {Ryans}, {Skillen}, {Street}, {Udry},
  and {Wheatley}]{Collier2007}
{Collier Cameron},~A. \latin{et~al.}  \emph{\mnras} \textbf{2007}, \emph{380},
  1230--1244\relax
\mciteBstWouldAddEndPuncttrue
\mciteSetBstMidEndSepPunct{\mcitedefaultmidpunct}
{\mcitedefaultendpunct}{\mcitedefaultseppunct}\relax
\EndOfBibitem
\bibitem[{Gillon} \latin{et~al.}(2017){Gillon}, {Triaud}, {Demory}, {Jehin},
  {Agol}, {Deck}, {Lederer}, {de Wit}, {Burdanov}, {Ingalls}, {Bolmont},
  {Leconte}, {Raymond}, {Selsis}, {Turbet}, {Barkaoui}, {Burgasser},
  {Burleigh}, {Carey}, {Chaushev}, {Copperwheat}, {Delrez}, {Fernandes},
  {Holdsworth}, {Kotze}, {Van Grootel}, {Almleaky}, {Benkhaldoun}, {Magain},
  and {Queloz}]{Gillon2017Natur}
{Gillon},~M. \latin{et~al.}  \emph{\nat} \textbf{2017}, \emph{542},
  456--460\relax
\mciteBstWouldAddEndPuncttrue
\mciteSetBstMidEndSepPunct{\mcitedefaultmidpunct}
{\mcitedefaultendpunct}{\mcitedefaultseppunct}\relax
\EndOfBibitem
\bibitem[{Gillon}(2018)]{Gillon2018}
{Gillon},~M. \emph{Nature Astronomy} \textbf{2018}, \emph{2}, 344--344\relax
\mciteBstWouldAddEndPuncttrue
\mciteSetBstMidEndSepPunct{\mcitedefaultmidpunct}
{\mcitedefaultendpunct}{\mcitedefaultseppunct}\relax
\EndOfBibitem
\bibitem[{Burdanov} \latin{et~al.}(2017){Burdanov}, {Delrez}, {Gillon},
  {Jehin}, {Speculoos}, and {Trappist Teams}]{Burdanov2017}
{Burdanov},~A.; {Delrez},~L.; {Gillon},~M.; {Jehin},~E.; {Speculoos},~T.;
  {Trappist Teams}, \emph{Handbook of Exoplanets, Edited by Hans J.~Deeg and
  Juan Antonio Belmonte.~Springer Living Reference Work, ISBN:
  978-3-319-30648-3, 2017, id.130}; 2017; p 130\relax
\mciteBstWouldAddEndPuncttrue
\mciteSetBstMidEndSepPunct{\mcitedefaultmidpunct}
{\mcitedefaultendpunct}{\mcitedefaultseppunct}\relax
\EndOfBibitem
\bibitem[{Delrez} \latin{et~al.}(2018){Delrez}, {Gillon}, {Queloz}, {Demory},
  {Almleaky}, {de Wit}, {Jehin}, {Triaud}, {Barkaoui}, {Burdanov}, {Burgasser},
  {Ducrot}, {McCormac}, {Murray}, {Silva Fernandes}, {Sohy}, {Thompson}, {Van
  Grootel}, {Alonso}, {Benkhaldoun}, and {Rebolo}]{Delrez2018}
{Delrez},~L. \latin{et~al.}  \emph{ArXiv e-prints} \textbf{2018}, \relax
\mciteBstWouldAddEndPunctfalse
\mciteSetBstMidEndSepPunct{\mcitedefaultmidpunct}
{}{\mcitedefaultseppunct}\relax
\EndOfBibitem
\bibitem[{Bonfils} \latin{et~al.}(2011){Bonfils}, {Gillon}, {Forveille},
  {Delfosse}, {Deming}, {Demory}, {Lovis}, {Mayor}, {Neves}, {Perrier},
  {Santos}, {Seager}, {Udry}, {Boisse}, and {Bonnefoy}]{Bonfils2011}
{Bonfils},~X.; {Gillon},~M.; {Forveille},~T.; {Delfosse},~X.; {Deming},~D.;
  {Demory},~B.-O.; {Lovis},~C.; {Mayor},~M.; {Neves},~V.; {Perrier},~C.;
  {Santos},~N.~C.; {Seager},~S.; {Udry},~S.; {Boisse},~I.; {Bonnefoy},~M.
  \emph{\aap} \textbf{2011}, \emph{528}, A111\relax
\mciteBstWouldAddEndPuncttrue
\mciteSetBstMidEndSepPunct{\mcitedefaultmidpunct}
{\mcitedefaultendpunct}{\mcitedefaultseppunct}\relax
\EndOfBibitem
\bibitem[{Gillon} \latin{et~al.}(2012){Gillon}, {Triaud}, {Fortney}, {Demory},
  {Jehin}, {Lendl}, {Magain}, {Kabath}, {Queloz}, {Alonso}, {Anderson},
  {Collier Cameron}, {Fumel}, {Hebb}, {Hellier}, {Lanotte}, {Maxted},
  {Mowlavi}, and {Smalley}]{Gillon2012}
{Gillon},~M. \latin{et~al.}  \emph{\aap} \textbf{2012}, \emph{542}, A4\relax
\mciteBstWouldAddEndPuncttrue
\mciteSetBstMidEndSepPunct{\mcitedefaultmidpunct}
{\mcitedefaultendpunct}{\mcitedefaultseppunct}\relax
\EndOfBibitem
\bibitem[{Delrez} \latin{et~al.}(2014){Delrez}, {Van Grootel}, {Anderson},
  {Collier-Cameron}, {Doyle}, {Fumel}, {Gillon}, {Hellier}, {Jehin}, {Lendl},
  {Neveu-VanMalle}, {Maxted}, {Pepe}, {Pollacco}, {Queloz}, {S{\'e}gransan},
  {Smalley}, {Smith}, {Southworth}, {Triaud}, {Udry}, and {West}]{Delrez2014}
{Delrez},~L. \latin{et~al.}  \emph{\aap} \textbf{2014}, \emph{563}, A143\relax
\mciteBstWouldAddEndPuncttrue
\mciteSetBstMidEndSepPunct{\mcitedefaultmidpunct}
{\mcitedefaultendpunct}{\mcitedefaultseppunct}\relax
\EndOfBibitem
\bibitem[{Jehin} \latin{et~al.}(2011){Jehin}, {Gillon}, {Queloz}, {Magain},
  {Manfroid}, {Chantry}, {Lendl}, {Hutsem{\'e}kers}, and {Udry}]{Jehin2011}
{Jehin},~E.; {Gillon},~M.; {Queloz},~D.; {Magain},~P.; {Manfroid},~J.;
  {Chantry},~V.; {Lendl},~M.; {Hutsem{\'e}kers},~D.; {Udry},~S. \emph{The
  Messenger} \textbf{2011}, \emph{145}, 2--6\relax
\mciteBstWouldAddEndPuncttrue
\mciteSetBstMidEndSepPunct{\mcitedefaultmidpunct}
{\mcitedefaultendpunct}{\mcitedefaultseppunct}\relax
\EndOfBibitem
\bibitem[{Gillon} \latin{et~al.}(2013){Gillon}, {Anderson}, {Collier-Cameron},
  {Doyle}, {Fumel}, {Hellier}, {Jehin}, {Lendl}, {Maxted}, {Montalb{\'a}n},
  {Pepe}, {Pollacco}, {Queloz}, {S{\'e}gransan}, {Smith}, {Smalley},
  {Southworth}, {Triaud}, {Udry}, and {West}]{Gillon2013}
{Gillon},~M. \latin{et~al.}  \emph{\aap} \textbf{2013}, \emph{552}, A82\relax
\mciteBstWouldAddEndPuncttrue
\mciteSetBstMidEndSepPunct{\mcitedefaultmidpunct}
{\mcitedefaultendpunct}{\mcitedefaultseppunct}\relax
\EndOfBibitem
\bibitem[Gillon \latin{et~al.}(2011)Gillon, Jehin, Magain, Chantry,
  Hutsem{\'{e}}kers, Manfroid, Queloz, and Udry]{Gillon2011}
Gillon,~M.; Jehin,~E.; Magain,~P.; Chantry,~V.; Hutsem{\'{e}}kers,~D.;
  Manfroid,~J.; Queloz,~D.; Udry,~S. \emph{{EPJ} Web of Conferences}
  \textbf{2011}, \emph{11}, 06002\relax
\mciteBstWouldAddEndPuncttrue
\mciteSetBstMidEndSepPunct{\mcitedefaultmidpunct}
{\mcitedefaultendpunct}{\mcitedefaultseppunct}\relax
\EndOfBibitem
\bibitem[Lendl \latin{et~al.}(2012)Lendl, Anderson, Collier-Cameron, Doyle,
  Gillon, Hellier, Jehin, Lister, Maxted, Pepe, Pollacco, Queloz, Smalley,
  S{\'{e}}gransan, Smith, Triaud, Udry, West, and Wheatley]{Lendl2012}
Lendl,~M. \latin{et~al.}  \emph{Astronomy {\&} Astrophysics} \textbf{2012},
  \emph{544}, A72\relax
\mciteBstWouldAddEndPuncttrue
\mciteSetBstMidEndSepPunct{\mcitedefaultmidpunct}
{\mcitedefaultendpunct}{\mcitedefaultseppunct}\relax
\EndOfBibitem
\bibitem[{Wheatley} \latin{et~al.}(2017){Wheatley}, {West}, {Goad}, {Jenkins},
  {Pollacco}, {Queloz}, {Rauer}, {Udry}, {Watson}, {Chazelas}, {Eigmuller},
  {Lambert}, {Genolet}, {McCormac}, {Walker}, {Armstrong}, {Bayliss}, {Bento},
  {Bouchy}, {Burleigh}, {Cabrera}, {Casewell}, {Chaushev}, {Chote},
  {Csizmadia}, {Erikson}, {Faedi}, {Foxell}, {Gansicke}, {Gillen}, {Grange},
  {Gunther}, {Hodgkin}, {Jackman}, {Jordan}, {Louden}, {Metrailler}, {Moyano},
  {Nielsen}, {Osborn}, {Poppenhaeger}, {Raddi}, {Raynard}, {Smith}, {Soto}, and
  {Titz-Weider}]{Wheatley2017}
{Wheatley},~P.~J. \latin{et~al.}  \emph{ArXiv e-prints} \textbf{2017}, \relax
\mciteBstWouldAddEndPunctfalse
\mciteSetBstMidEndSepPunct{\mcitedefaultmidpunct}
{}{\mcitedefaultseppunct}\relax
\EndOfBibitem
\bibitem[McCormac \latin{et~al.}(2014)McCormac, Skillen, Pollacco, Faedi,
  Ramsay, Dhillon, Todd, and Gonzalez]{McCormac2014}
McCormac,~J.; Skillen,~I.; Pollacco,~D.; Faedi,~F.; Ramsay,~G.; Dhillon,~V.~S.;
  Todd,~I.; Gonzalez,~A. \emph{Monthly Notices of the Royal Astronomical
  Society} \textbf{2014}, \emph{438}, 3383--3398\relax
\mciteBstWouldAddEndPuncttrue
\mciteSetBstMidEndSepPunct{\mcitedefaultmidpunct}
{\mcitedefaultendpunct}{\mcitedefaultseppunct}\relax
\EndOfBibitem
\bibitem[{Craig} \latin{et~al.}(2015){Craig}, {Crawford}, {Deil}, {Gomez},
  {G{\"u}nther}, {Heidt}, {Horton}, {Karr}, {Nelson}, {Ninan}, {Pattnaik},
  {Rol}, {Schoenell}, {Seifert}, {Singh}, {Sipocz}, {Stotts}, {Streicher},
  {Tollerud}, {Walker}, and {ccdproc contributors}]{Craig15}
{Craig},~M.~W. \latin{et~al.}  {ccdproc: CCD data reduction software}.
  Astrophysics Source Code Library, 2015\relax
\mciteBstWouldAddEndPuncttrue
\mciteSetBstMidEndSepPunct{\mcitedefaultmidpunct}
{\mcitedefaultendpunct}{\mcitedefaultseppunct}\relax
\EndOfBibitem
\bibitem[{Barbary}(2016)]{Barbary16}
{Barbary},~K. \emph{The Journal of Open Source Software} \textbf{2016},
  \emph{1}, 1\relax
\mciteBstWouldAddEndPuncttrue
\mciteSetBstMidEndSepPunct{\mcitedefaultmidpunct}
{\mcitedefaultendpunct}{\mcitedefaultseppunct}\relax
\EndOfBibitem
\bibitem[{Bertin} and {Arnouts}(1996){Bertin}, and {Arnouts}]{Bertin96}
{Bertin},~E.; {Arnouts},~S. \emph{\aaps} \textbf{1996}, \emph{117},
  393--404\relax
\mciteBstWouldAddEndPuncttrue
\mciteSetBstMidEndSepPunct{\mcitedefaultmidpunct}
{\mcitedefaultendpunct}{\mcitedefaultseppunct}\relax
\EndOfBibitem
\bibitem[{McCormac} \latin{et~al.}(2013){McCormac}, {Pollacco}, {Skillen},
  {Faedi}, {Todd}, and {Watson}]{McCormac2013}
{McCormac},~J.; {Pollacco},~D.; {Skillen},~I.; {Faedi},~F.; {Todd},~I.;
  {Watson},~C.~A. \emph{\pasp} \textbf{2013}, \emph{125}, 548\relax
\mciteBstWouldAddEndPuncttrue
\mciteSetBstMidEndSepPunct{\mcitedefaultmidpunct}
{\mcitedefaultendpunct}{\mcitedefaultseppunct}\relax
\EndOfBibitem
\bibitem[{Queloz} \latin{et~al.}(2000){Queloz}, {Mayor}, {Naef}, {Santos},
  {Udry}, {Burnet}, and {Confino}]{Queloz2000}
{Queloz},~D.; {Mayor},~M.; {Naef},~D.; {Santos},~N.; {Udry},~S.; {Burnet},~M.;
  {Confino},~B. \textbf{2000}, 548\relax
\mciteBstWouldAddEndPuncttrue
\mciteSetBstMidEndSepPunct{\mcitedefaultmidpunct}
{\mcitedefaultendpunct}{\mcitedefaultseppunct}\relax
\EndOfBibitem
\bibitem[{Baranne} \latin{et~al.}(1996){Baranne}, {Queloz}, {Mayor},
  {Adrianzyk}, {Knispel}, {Kohler}, {Lacroix}, {Meunier}, {Rimbaud}, and
  {Vin}]{Baranne1996}
{Baranne},~A.; {Queloz},~D.; {Mayor},~M.; {Adrianzyk},~G.; {Knispel},~G.;
  {Kohler},~D.; {Lacroix},~D.; {Meunier},~J.-P.; {Rimbaud},~G.; {Vin},~A.
  \emph{\aaps} \textbf{1996}, \emph{119}, 373--390\relax
\mciteBstWouldAddEndPuncttrue
\mciteSetBstMidEndSepPunct{\mcitedefaultmidpunct}
{\mcitedefaultendpunct}{\mcitedefaultseppunct}\relax
\EndOfBibitem
\bibitem[{Queloz} \latin{et~al.}(2001){Queloz}, {Henry}, {Sivan}, {Baliunas},
  {Beuzit}, {Donahue}, {Mayor}, {Naef}, {Perrier}, and {Udry}]{Queloz2001}
{Queloz},~D.; {Henry},~G.~W.; {Sivan},~J.~P.; {Baliunas},~S.~L.;
  {Beuzit},~J.~L.; {Donahue},~R.~A.; {Mayor},~M.; {Naef},~D.; {Perrier},~C.;
  {Udry},~S. \emph{\aap} \textbf{2001}, \emph{379}, 279--287\relax
\mciteBstWouldAddEndPuncttrue
\mciteSetBstMidEndSepPunct{\mcitedefaultmidpunct}
{\mcitedefaultendpunct}{\mcitedefaultseppunct}\relax
\EndOfBibitem
\bibitem[{Torres} \latin{et~al.}(2004){Torres}, {Konacki}, {Sasselov}, and
  {Jha}]{Torres2004}
{Torres},~G.; {Konacki},~M.; {Sasselov},~D.~D.; {Jha},~S. \emph{\apj}
  \textbf{2004}, \emph{614}, 979--989\relax
\mciteBstWouldAddEndPuncttrue
\mciteSetBstMidEndSepPunct{\mcitedefaultmidpunct}
{\mcitedefaultendpunct}{\mcitedefaultseppunct}\relax
\EndOfBibitem
\bibitem[{Doyle} \latin{et~al.}(2013){Doyle}, {Smalley}, {Maxted}, {Anderson},
  {Cameron}, {Gillon}, {Hellier}, {Pollacco}, {Queloz}, {Triaud}, and
  {West}]{Doyle2013}
{Doyle},~A.~P.; {Smalley},~B.; {Maxted},~P.~F.~L.; {Anderson},~D.~R.;
  {Cameron},~A.~C.; {Gillon},~M.; {Hellier},~C.; {Pollacco},~D.; {Queloz},~D.;
  {Triaud},~A.~H.~M.~J.; {West},~R.~G. \emph{\mnras} \textbf{2013}, \emph{428},
  3164--3172\relax
\mciteBstWouldAddEndPuncttrue
\mciteSetBstMidEndSepPunct{\mcitedefaultmidpunct}
{\mcitedefaultendpunct}{\mcitedefaultseppunct}\relax
\EndOfBibitem
\bibitem[{Doyle} \latin{et~al.}(2014){Doyle}, {Davies}, {Smalley}, {Chaplin},
  and {Elsworth}]{Doyle2014}
{Doyle},~A.~P.; {Davies},~G.~R.; {Smalley},~B.; {Chaplin},~W.~J.;
  {Elsworth},~Y. \emph{\mnras} \textbf{2014}, \emph{444}, 3592--3602\relax
\mciteBstWouldAddEndPuncttrue
\mciteSetBstMidEndSepPunct{\mcitedefaultmidpunct}
{\mcitedefaultendpunct}{\mcitedefaultseppunct}\relax
\EndOfBibitem
\bibitem[{Murray} and {Correia}(2010){Murray}, and {Correia}]{Murray2010}
{Murray},~C.~D.; {Correia},~A.~C.~M. In \emph{Exoplanets}; {Seager},~S., Ed.;
  2010; pp 15--23\relax
\mciteBstWouldAddEndPuncttrue
\mciteSetBstMidEndSepPunct{\mcitedefaultmidpunct}
{\mcitedefaultendpunct}{\mcitedefaultseppunct}\relax
\EndOfBibitem
\bibitem[{Mandel} and {Agol}(2002){Mandel}, and {Agol}]{Mandel2002}
{Mandel},~K.; {Agol},~E. \emph{\apjl} \textbf{2002}, \emph{580},
  L171--L175\relax
\mciteBstWouldAddEndPuncttrue
\mciteSetBstMidEndSepPunct{\mcitedefaultmidpunct}
{\mcitedefaultendpunct}{\mcitedefaultseppunct}\relax
\EndOfBibitem
\bibitem[Schwarz(1978)]{schwarz1978}
Schwarz,~G. \emph{Ann. Statist.} \textbf{1978}, \emph{6}, 461--464\relax
\mciteBstWouldAddEndPuncttrue
\mciteSetBstMidEndSepPunct{\mcitedefaultmidpunct}
{\mcitedefaultendpunct}{\mcitedefaultseppunct}\relax
\EndOfBibitem
\bibitem[{Claret}(2000)]{Claret2000}
{Claret},~A. \emph{VizieR Online Data Catalog} \textbf{2000}, \emph{336}\relax
\mciteBstWouldAddEndPuncttrue
\mciteSetBstMidEndSepPunct{\mcitedefaultmidpunct}
{\mcitedefaultendpunct}{\mcitedefaultseppunct}\relax
\EndOfBibitem
\bibitem[{Gelman} and {Rubin}(1992){Gelman}, and {Rubin}]{Gelman1992}
{Gelman},~A.; {Rubin},~D.~B. \emph{Statistical Science} \textbf{1992},
  \emph{7}, 457--472\relax
\mciteBstWouldAddEndPuncttrue
\mciteSetBstMidEndSepPunct{\mcitedefaultmidpunct}
{\mcitedefaultendpunct}{\mcitedefaultseppunct}\relax
\EndOfBibitem
\bibitem[{Enoch} \latin{et~al.}(2010){Enoch}, {Collier Cameron}, {Parley}, and
  {Hebb}]{Enoch2010}
{Enoch},~B.; {Collier Cameron},~A.; {Parley},~N.~R.; {Hebb},~L. \emph{\aap}
  \textbf{2010}, \emph{516}, A33\relax
\mciteBstWouldAddEndPuncttrue
\mciteSetBstMidEndSepPunct{\mcitedefaultmidpunct}
{\mcitedefaultendpunct}{\mcitedefaultseppunct}\relax
\EndOfBibitem
\bibitem[{Gaia Collaboration} \latin{et~al.}(2018){Gaia Collaboration},
  {Brown}, {Vallenari}, {Prusti}, {de Bruijne}, {Babusiaux}, {Bailer-Jones},
  {Biermann}, {Evans}, {Eyer}, and et~al.]{Gaia2018}
{Gaia Collaboration},; {Brown},~A.~G.~A.; {Vallenari},~A.; {Prusti},~T.; {de
  Bruijne},~J.~H.~J.; {Babusiaux},~C.; {Bailer-Jones},~C.~A.~L.;
  {Biermann},~M.; {Evans},~D.~W.; {Eyer},~L.; et~al., \emph{\aap}
  \textbf{2018}, \emph{616}, A1\relax
\mciteBstWouldAddEndPuncttrue
\mciteSetBstMidEndSepPunct{\mcitedefaultmidpunct}
{\mcitedefaultendpunct}{\mcitedefaultseppunct}\relax
\EndOfBibitem
\bibitem[{Pecaut} and {Mamajek}(2013){Pecaut}, and {Mamajek}]{Pecaut2013}
{Pecaut},~M.~J.; {Mamajek},~E.~E. \emph{\apjs} \textbf{2013}, \emph{208},
  9\relax
\mciteBstWouldAddEndPuncttrue
\mciteSetBstMidEndSepPunct{\mcitedefaultmidpunct}
{\mcitedefaultendpunct}{\mcitedefaultseppunct}\relax
\EndOfBibitem
\bibitem[{Maxted} \latin{et~al.}(2011){Maxted}, {Anderson}, {Collier Cameron},
  {Hellier}, {Queloz}, {Smalley}, {Street}, {Triaud}, {West}, {Gillon},
  {Lister}, {Pepe}, {Pollacco}, {S{\'e}gransan}, {Smith}, and
  {Udry}]{Maxted2011}
{Maxted},~P.~F.~L. \latin{et~al.}  \emph{\pasp} \textbf{2011}, \emph{123},
  547\relax
\mciteBstWouldAddEndPuncttrue
\mciteSetBstMidEndSepPunct{\mcitedefaultmidpunct}
{\mcitedefaultendpunct}{\mcitedefaultseppunct}\relax
\EndOfBibitem
\bibitem[{Maxted} \latin{et~al.}(2015){Maxted}, {Serenelli}, and
  {Southworth}]{Maxted2015}
{Maxted},~P.~F.~L.; {Serenelli},~A.~M.; {Southworth},~J. \emph{\aap}
  \textbf{2015}, \emph{575}, A36\relax
\mciteBstWouldAddEndPuncttrue
\mciteSetBstMidEndSepPunct{\mcitedefaultmidpunct}
{\mcitedefaultendpunct}{\mcitedefaultseppunct}\relax
\EndOfBibitem
\bibitem[{Weiss} and {Schlattl}(2008){Weiss}, and {Schlattl}]{Weiss2008}
{Weiss},~A.; {Schlattl},~H. \emph{\apss} \textbf{2008}, \emph{316},
  99--106\relax
\mciteBstWouldAddEndPuncttrue
\mciteSetBstMidEndSepPunct{\mcitedefaultmidpunct}
{\mcitedefaultendpunct}{\mcitedefaultseppunct}\relax
\EndOfBibitem
\bibitem[{Weiss} \latin{et~al.}(2013){Weiss}, {Marcy}, {Rowe}, {Howard},
  {Isaacson}, {Fortney}, {Miller}, {Demory}, {Fischer}, {Adams}, {Dupree},
  {Howell}, {Kolbl}, {Johnson}, {Horch}, {Everett}, {Fabrycky}, and
  {Seager}]{Weiss2013}
{Weiss},~L.~M. \latin{et~al.}  \emph{\apj} \textbf{2013}, \emph{768}, 14\relax
\mciteBstWouldAddEndPuncttrue
\mciteSetBstMidEndSepPunct{\mcitedefaultmidpunct}
{\mcitedefaultendpunct}{\mcitedefaultseppunct}\relax
\EndOfBibitem
\bibitem[{Ricker} \latin{et~al.}(2016){Ricker}, {Vanderspek}, {Winn}, {Seager},
  {Berta-Thompson}, {Levine}, {Villasenor}, {Latham}, {Charbonneau}, {Holman},
  {Johnson}, {Sasselov}, {Szentgyorgyi}, {Torres}, {Bakos}, {Brown},
  {Christensen-Dalsgaard}, {Kjeldsen}, {Clampin}, {Rinehart}, {Deming}, {Doty},
  {Dunham}, {Ida}, {Kawai}, {Sato}, {Jenkins}, {Lissauer}, {Jernigan},
  {Kaltenegger}, {Laughlin}, {Lin}, {McCullough}, {Narita}, {Pepper},
  {Stassun}, and {Udry}]{Richer2016TESS}
{Ricker},~G.~R. \latin{et~al.}  {The Transiting Exoplanet Survey Satellite}.
  Space Telescopes and Instrumentation 2016: Optical, Infrared, and Millimeter
  Wave. 2016; p 99042B\relax
\mciteBstWouldAddEndPuncttrue
\mciteSetBstMidEndSepPunct{\mcitedefaultmidpunct}
{\mcitedefaultendpunct}{\mcitedefaultseppunct}\relax
\EndOfBibitem
\end{mcitethebibliography}

\end{document}